\documentclass[10pt,aps,prd,twocolumn,superscriptaddress,amsmath,amssymb,nofootinbib]{revtex4-1}

\usepackage{graphicx}
\usepackage{dcolumn}
\usepackage{bm}

\usepackage[T1]{fontenc}
\usepackage{color,amsmath}
\usepackage{physics}
\usepackage{subfiles}
\usepackage{soul}
\usepackage{tabularray}
\usepackage{tensor}
\def\tns{\tensor}

\usepackage[normalem]{ulem}
\usepackage{colortbl}
\usepackage[usenames,dvipsnames,table,xcdraw]{xcolor}
\usepackage{hyperref}
\hypersetup{
    colorlinks,
    linkcolor={violet},
    citecolor={ForestGreen},
    urlcolor={teal}
}
\usepackage{orcidlink}

\newcommand{\beq}{\begin{eqnarray}}
\newcommand{\eeq}{\end{eqnarray}}
\newcommand{\bea}{\begin{eqnarray}}
\newcommand{\eea}{\end{eqnarray}}
\newcommand{\be}{\begin{equation}}
\newcommand{\ee}{\end{equation}}
\newcommand{\bq}{\begin{equation}}
\newcommand{\eq}{\end{equation}}

\newcommand{\half}{\frac{1}{2}}
\newcommand{\nn}{\nonumber}

\newcommand{\mpl}{M_{\rm Pl}}
\def\k{\kappa}

\def\l{\lambda}

\def\t{\tau}

\def\half{\frac12}

\def\m{\mu}
\def\n{\nu}

\def\d{\delta}
\def\6{\partial}

\def\a{\alpha}
\def\b{\beta}

\def\6{\partial}

\begin{document}

\title{Magnetic Anti-de Sitter Wormholes as seeds for Higgs Inflation}

\author{Panos Betzios\orcidlink{0000-0002-5350-9404}}
\email{panos.betzios@ugent.be}
\affiliation{\href{https://www.ugent.be/we/physics-astronomy/en}{Department of Physics and Astronomy},
Ghent University, \\ Krijgslaan, 281-S9, 9000 Gent, Belgium}

\author{Ioannis D.~Gialamas\orcidlink{0000-0002-2957-5276}}
\email{ioannis.gialamas@kbfi.ee}
 \affiliation{\href{https://kbfi.ee/high-energy-and-computational-physics/?lang=en}{Laboratory of High Energy and Computational Physics}, National Institute of Chemical Physics and Biophysics, \\
R{\"a}vala pst.~10, Tallinn, 10143, Estonia}

\author{Olga Papadoulaki\orcidlink{0000-0001-5302-2930}}
\email{olga.papadoulaki@polytechnique.edu}
\affiliation{\href{https://www.cpht.polytechnique.fr/?q=en}{CPHT, CNRS, École polytechnique, Institut Polytechnique de Paris}, \\ 91120 Palaiseau, France}

\date{\today}

\begin{abstract}
We show how certain types of magnetic asymptotically Anti-de Sitter Euclidean wormholes can catalyze the onset of inflation. These wormholes can be embedded as saddle point solutions of General Relativity coupled to the Standard Model, the inflaton being identified with the Higgs particle. Our scenario is based on the assumption that the quantum effective potential for the Higgs turns negative at a certain high energy window, in line with current measured values for the Higgs and Top quark masses. Within our proposal, we can estimate various parameters and physical quantities of interest, in consistency with current observational bounds.
\end{abstract}

\maketitle


\section{Introduction}

The tremendous advances of theoretical and experimental physics during the previous century led to an extremely detailed understanding of the history and evolution of our Universe, from the present era back to its primordial origin, when its size was billion times smaller than that of a proton. Cosmic inflation~\cite{Kazanas:1980tx,Guth:1980zm,Linde:1981mu} is our leading candidate theory describing the physics
of the rapidly expanding early universe at this primordial epoch, leading to predictions that conform very well with our current cosmological observational data - a surprising feat considering the vast hierarchy of scales between the primordial and current epoch.

Nevertheless the onset of inflation and the epochs preceding it were never really understood in detail, despite numerous bold efforts of theoretical physicists such as~\cite{Hartle:1983ai,Vilenkin:1983xq}. Effective field theory itself offers an operationally concrete setting to pose the relevant questions: Why and how did the universe start with appropriate conditions high up on the inflaton potential? While these being initial/boundary value questions  within the context of purely inflationary theory, we do expect that a more complete theory could explain them in a meaningful and natural way, and perhaps in a more dynamical fashion.

In this paper we \emph{propose and show how certain types of semi-classical solutions of General Relativity coupled to the Standard Model can catalyze the onset of inflation}. These are $Z_2$ ($\tau \leftrightarrow - \tau$) symmetric ``wineglass anti-de Sitter wormholes''~\cite{Betzios:2024oli} that describe the non-perturbative nucleation process of the inflationary universe, whose initial state is determined from the $\tau = 0$ slice of the Euclidean wormhole geometry.

Our paradigm is based on two assumptions: that the inflaton is to be identified with the experimentally observed Higgs particle
(the so called ``Higgs inflation'' scenario~\cite{Bezrukov:2007ep}), and that the effective Higgs potential turns negative at a certain high energy window~\cite{Bezrukov:2012sa,Buttazzo:2013uya}. While the first assumption is not entirely crucial for the viability of our scenario, it naturally resolves the problem of explaining what happens to the inflaton in our current energy scales and why we have not yet observed any other scalar particle besides the Higgs. The second assumption while being structurally more important for our proposal, it aligns with current experimental data for the Higgs and Top quark masses~\cite{ParticleDataGroup:2024cfk} that lead to a metastability of the electroweak vacuum~\cite{Buttazzo:2013uya}. 

Our proposed model has various phenomenologically interesting outputs. 
As other models of Higgs inflation (that essentially generalise Starobinsky's $R^2$ model~\cite{Starobinsky:1980te}), it can explain very well current cosmological data~\cite{Planck:2018jri,BICEP:2021xfz}.  At the same time it fares better regarding some troublesome points of former Higgs inflation models such as the unitarity issue~\cite{ Barbon:2009ya,Bezrukov:2010jz}. It also predicts the presence of primordial magnetic fields with a very large coherence length. While these magnetic fields get diluted to an enormous degree due to the accelerated inflationary expansion~\cite{Turner:1987bw}, they could perhaps play an important role even at later stages of the evolution of the universe if further physical mechanisms that can amplify and sustain them become active~\cite{Hogan:1983zz,Turner:1987bw,Kandus:2010nw,Durrer:2013pga,Subramanian:2015lua}. Finally, it has the potential of being embeddable in a UV complete model of quantum gravity such as string theory, due to the Euclidean AdS asymptotics of the relevant wormhole backgrounds~\cite{Maldacena:2004rf,Betzios:2019rds,Betzios:2021fnm,VanRaamsdonk:2021qgv,Antonini:2022ptt,Betzios:2024oli}.

We use the abbreviations: (SM) for the Standard Model, (EW) for the Electroweak scale, (EFT) for Effective Field Theory (RG) for the Renormalization Group, (AdS) for Anti-de Sitter and natural units $\hbar = c = k_B =\epsilon_0 =1$ throughout.

\section{The Model}

Our starting point is the EFT composed of the Einstein-Hilbert action together with a nonminimally coupled scalar field $\phi$ (that will be identified with the Higgs field) and the rest of the radiation and matter fields of the SM. In Euclidean signature the action is
\begin{align}
\label{eq:act_1}
\mathcal{S}_E = \int {\rm d}^4x \sqrt{g_E} \bigg[ &-\frac{1}{2\kappa}R - \frac{\xi \phi^2}{2} R +\frac12 \partial_\m \phi \partial^\m \phi +V(\phi) \nonumber
\\
&+ \mathcal{L}^{E}_{\rm rad} \, + \, \mathcal{L}^{E}_{\rm matter} \bigg]\,,
\end{align}
where $\kappa \equiv 1/\mpl^2$ and $V(\phi)$ is the scalar field potential. The non-minimal coupling term typically arises, when considering the one-loop effective action of the SM coupled to gravity via quantum loop corrections and is required to renormalise the stress energy tensor of scalar fields in curved backgrounds~\cite{Birrell:1982ix}. The scalar potential, should then also be understood to 
be the loop-corrected effective potential. In general, the EFT contains higher derivative and more complicated non-minimal coupling terms,  such as $W(\phi) R^m$. In this work we focus in the \emph{minimal Higgs inflation} scenario where the $\xi \phi^2 R$ term is the most relevant term in such an expansion~\cite{Bezrukov:2007ep,Bezrukov:2010jz}. This is further corroborated by the fact that the characteristic size of the throat of our wormhole backgrounds, is at least $ 10^{5} \mpl^{-1}$, rendering higher curvature and quantum corrections subdominant~\footnote{See the \hyperlink{sup_mat}{\color{violet}Supplementary Material} for details.}.

The action~\eqref{eq:act_1} can be expressed in the Einstein frame (EF) through a Weyl rescaling of the form
\be
\label{eq:Weyl}
g_{\m\n} \rightarrow \Omega^{-2}(\phi) g_{\m\n}\,, \qquad \text{with} \qquad \Omega^{2}(\phi) = 1 +\kappa\xi\phi^2\,.
\ee
After performing a field redefinition of the form
\be
{\rm d}\chi = \sqrt{\left(1+(1+6\xi)\kappa\xi\phi^2\right)/(1+\kappa\xi\phi^2)^2}\, {\rm d}\phi\,,
\ee
the Einstein frame action reads,
\be
\mathcal{S}_E = \int {\rm d}^4x\sqrt{g_E} \left(-\frac{R}{2\kappa} +\frac{\partial_\m\chi\partial^\m\chi}{2} +U(\chi) + \mathcal{L}^{E}_{\rm rad, matter}   \right)\,,
\ee
where the potential of the canonical scalar field is given by $U(\chi) = V(\phi(\chi))/ \Omega^{4}(\phi(\chi))$.

Since we want to understand the dynamics of the early universe (having a small size), the leading contribution to the Einstein equations stems from the stress energy tensor of possible axionic fields scaling as $a^{-6}$, followed by that of the radiation and (non-relativistic) matter fields, scaling as $a^{-4}$ and $a^{-3}$ respectively ($a$ is the scale factor in the FLRW ansatze~\eqref{eq:ansatze}). The axionic contribution was analysed in~\cite{Betzios:2024oli}, but in what follows, we shall only consider the radiation (density and pressure) components to the stress energy tensor $\rho_{\rm rad}, P_{\rm rad}$, arising from the experimentally observed SM gauge fields~\footnote{We also include
in $\rho_{\rm rad}$ the contribution of any relativistic matter such as neutrinos. To first approximation we neglect any non-relativistic matter contribution since it is subdominant.}. Another advantage is that their actions are scale-invariant and so their contribution is also invariant under the rescaling~\eqref{eq:Weyl}.

Our approach is phenomenological, as we do not solve the resulting equations of motion (EOMs) for the radiation fields directly, but treat them as a fluid (see~\cite{Rey:1989th,Hosoya:1989zn,Betzios:2019rds,Marolf:2021kjc,Deshpande:2022zfm,Lan:2024gnv} for wormhole examples with radiation fields where the complete field equations are solved). However, this approach suffices to demonstrate the existence of the backgrounds we are interested in and has the merit that it applies to radiation fields in a universal way -- all the SM radiation fields contribute to our background solutions. While in their Lorentzian equation of state
\be
  P_{\rm rad}^L = \frac{1}{3} \rho_{\rm rad}^L \, , \qquad \rho_{\rm rad}^L \geq 0 \, ,
\ee
their energy density is positive definite, the ``Euclidean energy'' density of the radiation, $\rho_{\rm rad}^E$, can be negative if the field has a dominant magnetic component. For example, in the $U(1)$ E/M case
the ``energy'' density from the Euclidean stress energy tensor is~\cite{Gupta:1989bs}
\be
T^E_{00}  = \half \left(  E^2 - B^2 \right)\,.
\ee
We therefore find that the magnetic component of the gauge fields gives rise to a negative contribution to the Euclidean ``energy'', while the electric component to a positive one (a similar result holds for the axion written as a two-form field).
As we demonstrate, \emph{a negative Euclidean energy density is essential to our scenario, implying the necessity of a magnetic radiation component} that was prominent in the early universe.
\begin{figure}[t!]
\centering
\includegraphics[width=0.49\textwidth]{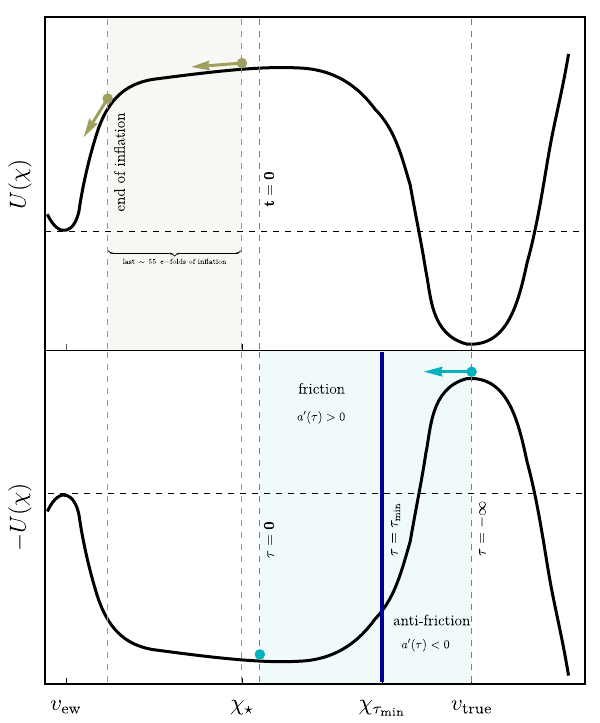}
\caption{{\textbf{Upper:}} The Lorentzian evolution. In this case the scalar field naturally reaches the EW vacuum. The condition of having a sufficiently large number of e-folds (wide hilltop region) also helps in the stability of the EW vaccuum.  {\textbf{Lower:}} The Euclidean evolution corresponding to a half wormhole configuration. It starts with anti-friction $a'(\tau)<0$, since the space is asymptotically EAdS. At $\tau = 0 = t$ the universe nucleates and then follows the slow-roll trajectory.}
\label{fig:1}
\end{figure}

Arguments regarding the dominance of $SO(4)$ symmetric Euclidean configurations in the path integral~\footnote{See for example the lectures \cite{Coleman:1985rnk}.}, consistent with the observed isotropy and homogeneity of our universe, lead us to use the following Euclidean FLRW type of ansatze for describing its nucleation process 
\be\label{eq:ansatze}
{\rm d} s^2_{\rm EFLRW} =  {\rm d} \t^2 +a^2(\t)  {\rm d} \Omega_3^ 2\, , \quad \chi =\chi(\t) \, .
\ee
In this ansatze, the Euclidean EOMs take the form~\footnote{{To simplify notation, we define the tilded quantities $\tilde{U} = \k U/3$, $\tilde{\chi} = \sqrt{\k/3}\chi$ and $ \kappa  \rho_{\rm rad}^E/3 = \tilde{\rho}_{\rm rad}^E /a^4 $. The subscripts $``0"$ and $``\tau_{\rm min}"$ indicate values calculated at the local maximum and global minimum of the scale factor, respectively.}} 
\begin{align}
\tilde{\chi}''+3\frac{a'\tilde{\chi}'}{a}-\frac{{\rm d}\tilde{U}(\chi)}{{\rm d} \tilde{\chi}} &= 0\,, \nonumber
\\
\frac{2a''}{a}+\frac{a'^2}{a^2}-\frac{1}{a^2}  + 3 \left(\tilde{U}(\chi)+\frac{\tilde{\chi}'^2}{2} \right) +  \frac{\tilde{\rho}_{\rm rad}^E}{a^4} & =0  \,, \nonumber
\\ 
\frac{a'^2}{a^2} -\frac{1}{a^2}  + \left(\tilde{U}(\chi)-\frac{\tilde{\chi}'^2}{2} \right) - \frac{\tilde{\rho}_{\rm rad}^E}{a^4} & =0 \,, 
\label{eq:eoms}
\end{align}
where the prime denotes a derivative with respect to $\tau$. The conditions that describe a ``wineglass'' shaped wormhole with a local maximum $a(0) = a_{\rm max}$ for the scale factor are~\cite{Betzios:2024oli}
\be
a''(0)<0\,, \quad a'(0) = 0\,,  \quad \chi'(0) = 0\,. 
\ee
These are favorable initial conditions for a subsequent Lorentzian inflationary evolution after the universe nucleates at $\tau = 0$~\cite{Betzios:2024oli}, since upon $\tau = i t , \,$ ${\rm d}^2 a/{\rm d}t^2 (0) >0  $. Additionally, there is another point in the Euclidean evolution where the scale factor must reach a local minimum, $a_{\rm min}$, i.e. $a'(\tau_{\rm min}) = 0$ and $a''(\tau_{\rm min})>0$. Since the Euclidean evolution of the scalar happens in the potential $- \tilde{U}(\chi)$~\cite{Betzios:2024oli}, in order to find the wormhole solutions of our interest, we assume an effective potential $\tilde{U}(\chi)$ with a positive global maximum and a negative global minimum, for which (see fig.~\ref{fig:1} for an example)
\be\label{eq:potineq}
\tilde{U}_{\text{global max.}} > \tilde{U}_0 > \tilde{U}_{\tau_{\rm min}} >  \tilde{U}_{\text{global min.}} \, .
\ee
From the third equation in~\eqref{eq:eoms}, we deduce that when $a'=0$, $x\equiv a^2 $ obeys a quadratic equation with solutions
\be
x_{\pm} = \frac{-1 \mp \sqrt{1 - 4 \mathcal{H}_E \tilde{\rho}_{\rm rad}^E}}{2 \mathcal{H}_E } \, , \quad \mathcal{H}_E = \frac{\tilde{\chi}'^2}{2} - \tilde{U}(\chi) \, ,
\label{eq:solx}
\ee
where $ \mathcal{H}_E \geq - \tilde{U}_0$ plays the role of a ``Euclidean energy'', that eventually gets dissipated during the motion in the lower fig.~\ref{fig:1}. Combining the second and third equations in~\eqref{eq:eoms}, we determine the acceleration
when $a'=0$
\be
\frac{a''}{a} = - 3 \tilde{U} -\mathcal{H}_E + \frac{1}{x_\pm}  \, .
\label{eq:add_0}
\ee
A short analysis shows that it is impossible to satisfy the wormhole conditions when $\tilde{\rho}_{\rm rad}^E > 0$, so a negative energy density is needed. This requirement arises from a Euclidean analog of gravitational focusing where gravity shrinks the space/scale factor, while negative Euclidean energy (magnetic flux in our case) enables a bounce and re-expansion. The competition is between the curvature term $1/a^2 $ and the flux term $ \sim 1/a^4 $, which dominates at small \( a \). Other fields, like axions ($\sim 1/a^6 $) or magnetized matter ($ \sim 1/a^3 $), can play a similar role.

\begin{figure}[t!]
\centering
\includegraphics[width=0.49\textwidth]{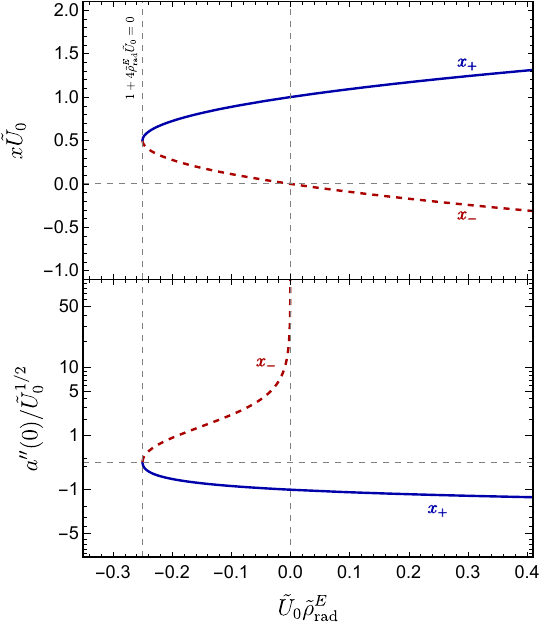}
\caption{The normalized scale factor and its second derivative at $\tau=0$ as functions of $\tilde{U}_0\tilde{\rho}_{\rm rad}^E $. The conditions $a_{\rm max}>0$, and $a''(0)<0$ are satisfied for $0>\tilde{U}_0 \tilde{\rho}_{\rm rad}^E  > -\frac14$ when using the $x_+$ branch.}
\label{fig:2}
\end{figure}

Let us now assume that $\tilde{\rho}_{\rm rad}^E < 0$, with
\be\label{eq:radbound}
0 \, > \, \tilde{U}_0  \tilde{\rho}_{\rm rad}^E \, \geq \, - \frac{1}{4} \, ,
\ee 
so that the scale factor is real. We now find
both branches of solutions $x_\pm$, $x_+ \geq x_-$, becoming the same at the boundary of the inequality above, see fig.~\ref{fig:2}. We therefore choose $a_{\tau_{\rm min}}$ in the $x_-$  and $a_{max}$ in the $x_+$ branch. For the $a_{max}$ solution ($\mathcal{H}_E = - \tilde{U}_0$) we find that
$a''(0) < 0 $ always. For the $x_-$ solution, we find that
$a''(\tau_{\rm min}) > 0$ as long as $\mathcal{H}_E^{\rm max} > \mathcal{H}_E(\tau_{\rm min})  > - \tilde{U}_0$. The existence of the upper bound is tied to the fact that if $\mathcal{H}_E$ is too high at $\tau_{\rm min}$ (for example if one sends $\tilde{U}_{\text{global min.}} \rightarrow -\infty$ in fig.~\ref{fig:1}), the friction will not be enough to stop the particle, resulting in overshooting. In the rest we shall assume that the effective potential does not exhibit such pathological characteristics~\footnote{Another possibility is that the potential in the Euclidean region has both a local minimum and maximum as studied in~\cite{Betzios:2024oli}, so that the Euclidean motion in the inverted potential cannot exhibit overshooting.}.

The result of this analysis is that we can generically obtain the wanted wormhole solutions for a spatially compact early universe, using relativistic fields with a slightly dominant magnetic component, as long as the inequality \eqref{eq:radbound} is respected.
In this paper we do not analyse further possible mechanisms by which such primordial magnetic fields can be produced, see~\cite{Giovannini:2003yn,Brandenburg:2004jv,Barrow:2006ch,Subramanian:2015lua,Kandus:2010nw} for reviews that focus on their generation during and after inflation.

\section{Higgs inflation}
As discussed earlier, our model can be embedded in Higgs inflation~\cite{Bezrukov:2007ep}, if we consider a tree level potential $V(\phi) = \lambda (\phi^2 - v_{\rm ew}^2)^2/4  $, where quantum loop effects induce a running quartic coupling, $\lambda(\mu)$ ($\mu$ is the renormalization group (RG) scale), that remains positive up to the inflationary scale $\mu_\star$. The non-minimal coupling, $\xi(\mu)$, has the effect of flattening the effective potential at large field values in the Einstein frame, which can lead to  a sufficient number of $e-$ folds for inflation. Assuming values for the Higgs and Top quark mass around the central values provided by~\cite{ParticleDataGroup:2024cfk}, we seem to be in the metastable region for the EW vacuum. This introduces a negative minimum ($v_{\rm true}$) in the effective potential, and the quartic coupling turns negative in a certain energy window~\footnote{If the Higgs mass, the SU(3) gauge coupling, and other measured SM parameters are fixed at their central values, there is always a critical top quark mass for which the electroweak and Planck-scale vacua become degenerate (see, e.g.~\cite{Masina:2024ybn}). Below this threshold, the Higgs effective potential remains stable, while for values above it, becomes metastable. Experimental data place us in the metastable region~\cite{ParticleDataGroup:2024cfk}, supporting our proposal. While other parameters ($m_h, g_3, M_Z$, etc.) affect stability, a full analysis of the parameter space is beyond the scope of this paper.}. We are interested in the parametric regime which places this window above the inflationary scale $\mu_\star$, corresponding to earlier times in the evolution of the universe prior to the onset of inflation~\footnote{In~\cite{Bezrukov:2014ipa}, an alternative scenario was considered in which the true (negative) minimum lies between the electroweak vacuum and the plateau region and hence affects the Lorentzian time evolution. This scenario is not very satisfactory, since unless one invokes certain thermal effects that completely lift this minimum, inflation will end there and the resulting cosmology will be in stark contrast with observations.}. In this scenario, we obtain an inflationary hilltop plateau for $v_{\rm true}>\chi>v_{\rm ew}$, and the shape of the quantum corrected effective potential is shown schematically in the upper panel of Figure~\ref{fig:1}. While we obtain this scenario for appropriate values of SM parameters, well within experimental bounds (see~\cite{Fumagalli:2016lls,Enckell:2018kkc,Masina:2024ybn} for more detailed discussions and a list of references), we remain agnostic regarding the actual value of $v_{\rm true}$ and the detailed shape of the effective potential at such high energy scales, since it is sensitive to the precise UV-completion of our model and the scale at which beyond the SM physics, such as supersymmetry,
could kick in. Nevertheless in low energy effective actions arising from string theory, one generically finds effective potentials with a global negative minimum, typically preserving some amount of supersymmetry, so our scenario is very natural in potential  string motivated UV completions of our model~\footnote{In this case, one finds that the opposite is harder to achieve - that is to uplift a part of the effective scalar potential to positive values in a controllable way, see~\cite{Baumann:2014nda} for a review.}.  

If we adopt our scenario of Higgs inflation, the potential in the plateau region at horizon crossing is approximately $U(\chi_\star) \simeq \lambda_\star/ (4\kappa^2 \xi^2_\star)$ and this ratio of the couplings remains approximately constant for most of the inflationary period due to the slow-roll condition. The amplitude of the primordial curvature perturbation on super-Hubble scales generated by single-field slow-roll inflation, $A^\star_s =  \frac{N_\star^2 \lambda_\star}{72 \pi^2 \xi^2_\star}$, is constrained by recent observations~\cite{Planck:2018jri} to approximately  $\sim 2.1\times 10^{-9}$ at the pivot scale $k_\star = 0.05$ Mpc$^{-1}$.  This constraint, in turn, determines the ratio $\lambda_\star/\xi^2_\star \sim (4-5)\times 10^{-10}$ and $U(\chi_\star) \simeq (1.0-1.2)\times 10^{-10} \,\mpl^4$, for $N_\star = 60-55$ $e$-folds of inflation.

We should emphasize at this point that these conditions are at the inflationary scale, at lower energies from the end of inflation to the current EW scale $\lambda(\mu)$ has an appreciable non-trivial running, so that~\cite{Buttazzo:2013uya} $\lambda_{\rm ew} \simeq 0.126 \gg \lambda_\star  $ (above the inflationary scale it turns negative). Hence our model accommodates a non-minimal coupling whose value $\xi_\star$ at the inflationary scale is much lower than that of the standard tree level discussions of Higgs inflation, where $\lambda_\star \simeq \lambda_{\rm ew} $~\cite{Bezrukov:2007ep,Rubio:2018ogq}. This result helps with the issue of violating perturbative unitarity at the scale $\Lambda_u = \mpl/\xi$, that Higgs inflation models exhibit~\cite{Burgess:2009ea, Barbon:2009ya, Burgess:2010zq, Bezrukov:2010jz}, since for small $\xi_\star$ it is pushed well above the inflationary scale, and towards the Planck scale, where we inevitably need to assume a certain UV-completion of our EFT model.

Within our model, we can also make an estimate on the value of primordial magnetic fields at the end of inflation and compare it with current experimental bounds. At the time $\tau = t =0$, where the Lorentzian evolution begins, the field is already positioned in the plateau region, allowing us to approximate
\be
U_0 \simeq U(\chi_\star) \simeq (1.0-1.2)\times 10^{-10} \,\mpl^4 \, , 
\label{eq:Uvalue}
\ee
Using eqns.~\eqref{eq:solx} and~\eqref{eq:radbound} we find that the maximum value of the scale factor is approximately $a^2_{\rm max} \simeq {3}/{(\kappa U_0)}$. Substituting this in the definition $ \kappa  \rho_{\rm rad}^E/3 = \tilde{\rho}_{\rm rad}^E /a^4 $ and using eq.~\eqref{eq:Uvalue} we obtain
\be
{\rho_{\rm rad}^E}_\star \simeq \frac{\kappa {U}_0^2 \tilde{\rho}^E_{\rm rad}}{3} \, ,  \quad 0 >{\rho_{\rm rad}^E}_\star > - (2.5-3.0) \times 10^{-11} \mpl^4 \,.
\ee
We then separate from this density its magnetic component ${\rho_{\rm rad}^E}_\star = {\rho_{\rm rest}^E}_\star - B^2_\star/2$. Assuming that no fine-tuning takes place, but rather a small asymmetry between the magnitudes of the magnetic and electric components, the scale of the magnetic field density at the beginning of inflation is $B^2_\star \simeq 10^{-11} \mpl^4 $.  During inflation, the universe was a very poor conductor, leading to an adiabatic dilution of this primordial density by a factor of $\simeq 10^{-96} - 10^{-104}$ for $N_\star = 55-60$ $e$-folds, resulting into $B_{\rm reh} \simeq 1000 - 0.1 \, {\rm G} $  at the beginning of reheating. The rapid expansion also leads to fields with very large coherence length after horizon re-entry. During reheating and up until horizon re-entry the universe turns into a very good conductor. By the time we enter the radiation era, the electric fields get eliminated by the currents of the highly conducting plasma, and the universe is left with classical large-scale magnetic fields~\cite{Turner:1987bw}. One
can then define a ratio $r_B = {B^2}/8\pi{\rho_{\gamma}} 
$ ($\rho_\gamma$ is the energy density of relativistic particle species), that remains approximately constant during the history of the universe from reheating until galaxy formation~\cite{Turner:1987bw}. Possible galactic dynamo mechanisms~\cite{Kulsrud:1992rk,Beck:1995zs,Kulsrud:1999bg,Widrow:2002ud,Brandenburg:2004jv} that can sustain the currently observed magnetic fields require a minimum ``seed'' field of the order of $B_{\rm seed} \simeq  10^{-22} $ G (that is model dependent) and a coherence scale of at least $10$ Kpc. Since $\rho_\gamma \simeq 10^{-51} \,{\rm GeV}^4$ today, we find a sufficient seed ratio $r_B \simeq 10^{-34}$. Evolving back the current radiation fields to the reheating era, we find $\rho_{\gamma , {\rm reh}} \simeq 10^{62}  \,{\rm GeV}^4$ assuming instantaneous thermalization.
In consistency with~\cite{Turner:1987bw} we find that the magnetic fields in our model after reheating, while being coherent to extremely large scales, they get diluted to such a high degree during the expansion of the universe, making it impossible for them to seed the galactic dynamo mechanism
unless additional physical mechanisms are invoked~\cite{Hogan:1983zz,Turner:1987bw,Kandus:2010nw,Subramanian:2015lua}.

\section{Discussion}

In this paper, we proposed that certain magnetic Euclidean (half) wormhole saddles can catalyze the onset of inflation, while at the same time being embeddable in a model of Higgs inflation containing only experimentally observed SM fields. This constitutes an important development compared to~\cite{Betzios:2024oli}, since the models we consider in the present work lead to predictions that are in line with current observational data. 

The most interesting future direction is to analyze in great detail the quantum effective potential for the Higgs, assuming various scenarios for particle physics at the inflationary scale and above (for example the presence or not of Supersymmetry and Grand-Unification). Our wormhole backgrounds exist only if the electroweak vacuum is metastable, the energy window where the effective potential turns negative being higher than the inflationary scale. RG corrections, could also potentially leave intermediate scale imprints affecting the detailed numbers of observational quantities, such as the spectral index. This allows us to place stringent bounds on particle masses and perhaps other SM parameters, by solving the RG equations for each such scenario and therefore to solidify or rule out our proposal.
The presence of primordial magnetic fields in our model also offers a potential window into understanding the origin of the large scale magnetization of our Universe, albeit it suffers from the same inflationary dilution issues that plague similar proposals in the literature.

Finally, due to their Anti-de Sitter asymptotics, such Euclidean wormhole saddles~\cite{Maldacena:2004rf,Betzios:2019rds,Betzios:2021fnm,VanRaamsdonk:2021qgv,Antonini:2022ptt,Betzios:2024oli} could descend from a fundamental UV complete construction (String Theory or the AdS/CFT correspondence). At a more conservative level, these Euclidean saddles are an effective proxy for the expected complicated chaotic dynamics of the early universe, allowing us to perform consistent semi-classical computations, in analogy to studies of quantum mechanical black holes~\cite{Almheiri:2020cfm}.

\begin{acknowledgments}

We wish to acknowledge useful discussions with several colleagues in conferences and seminars where we have presented our work. We also wish to thank the anonymous referees for their comments that helped us improve the quality and clarity of the manuscript.
\\
\noindent P.B. acknowledges financial support from the European Research Council (grant BHHQG-101040024), funded by the European Union.
P.B. also acknowledges support from the Natural Sciences and Engineering Research Council of Canada and the Simons foundation, at the initial stages of this project, when he was a member of the University of British Columbia. \\
The work of IDG was supported by the Estonian Research Council grants MOB3JD1202, RVTT3,  RVTT7, and by the CoE program TK202 ``Fundamental Universe''. \\
Views and opinions expressed are those of the authors only and do not necessarily reflect those of the European Union or the European Research Council. Neither the European Union nor the granting authority can be held responsible for them.

\end{acknowledgments}

\bibliography{biblio}

\providecommand{\noopsort}[1]{}\providecommand{\singleletter}[1]{#1}%
\begin{thebibliography}{77}%
\makeatletter
\providecommand \@ifxundefined [1]{%
 \@ifx{#1\undefined}
}%
\providecommand \@ifnum [1]{%
 \ifnum #1\expandafter \@firstoftwo
 \else \expandafter \@secondoftwo
 \fi
}%
\providecommand \@ifx [1]{%
 \ifx #1\expandafter \@firstoftwo
 \else \expandafter \@secondoftwo
 \fi
}%
\providecommand \natexlab [1]{#1}%
\providecommand \enquote  [1]{``#1''}%
\providecommand \bibnamefont  [1]{#1}%
\providecommand \bibfnamefont [1]{#1}%
\providecommand \citenamefont [1]{#1}%
\providecommand \href@noop [0]{\@secondoftwo}%
\providecommand \href [0]{\begingroup \@sanitize@url \@href}%
\providecommand \@href[1]{\@@startlink{#1}\@@href}%
\providecommand \@@href[1]{\endgroup#1\@@endlink}%
\providecommand \@sanitize@url [0]{\catcode `\\12\catcode `\$12\catcode `\&12\catcode `\#12\catcode `\^12\catcode `\_12\catcode `\%12\relax}%
\providecommand \@@startlink[1]{}%
\providecommand \@@endlink[0]{}%
\providecommand \url  [0]{\begingroup\@sanitize@url \@url }%
\providecommand \@url [1]{\endgroup\@href {#1}{\urlprefix }}%
\providecommand \urlprefix  [0]{URL }%
\providecommand \Eprint [0]{\href }%
\providecommand \doibase [0]{http://dx.doi.org/}%
\providecommand \selectlanguage [0]{\@gobble}%
\providecommand \bibinfo  [0]{\@secondoftwo}%
\providecommand \bibfield  [0]{\@secondoftwo}%
\providecommand \translation [1]{[#1]}%
\providecommand \BibitemOpen [0]{}%
\providecommand \bibitemStop [0]{}%
\providecommand \bibitemNoStop [0]{.\EOS\space}%
\providecommand \EOS [0]{\spacefactor3000\relax}%
\providecommand \BibitemShut  [1]{\csname bibitem#1\endcsname}%
\let\auto@bib@innerbib\@empty
\bibitem [{\citenamefont {Kazanas}(1980)}]{Kazanas:1980tx}%
  \BibitemOpen
  \bibfield  {author} {\bibinfo {author} {\bibfnamefont {D.}~\bibnamefont {Kazanas}},\ }\href {\doibase 10.1086/183361} {\bibfield  {journal} {\bibinfo  {journal} {Astrophys. J. Lett.}\ }\textbf {\bibinfo {volume} {241}},\ \bibinfo {pages} {L59} (\bibinfo {year} {1980})}\BibitemShut {NoStop}%
\bibitem [{\citenamefont {Guth}(1981)}]{Guth:1980zm}%
  \BibitemOpen
  \bibfield  {author} {\bibinfo {author} {\bibfnamefont {A.~H.}\ \bibnamefont {Guth}},\ }\href {\doibase 10.1103/PhysRevD.23.347} {\bibfield  {journal} {\bibinfo  {journal} {Phys. Rev. D}\ }\textbf {\bibinfo {volume} {23}},\ \bibinfo {pages} {347} (\bibinfo {year} {1981})}\BibitemShut {NoStop}%
\bibitem [{\citenamefont {Linde}(1982)}]{Linde:1981mu}%
  \BibitemOpen
  \bibfield  {author} {\bibinfo {author} {\bibfnamefont {A.~D.}\ \bibnamefont {Linde}},\ }\href {\doibase 10.1016/0370-2693(82)91219-9} {\bibfield  {journal} {\bibinfo  {journal} {Phys. Lett. B}\ }\textbf {\bibinfo {volume} {108}},\ \bibinfo {pages} {389} (\bibinfo {year} {1982})}\BibitemShut {NoStop}%
\bibitem [{\citenamefont {Hartle}\ and\ \citenamefont {Hawking}(1983)}]{Hartle:1983ai}%
  \BibitemOpen
  \bibfield  {author} {\bibinfo {author} {\bibfnamefont {J.~B.}\ \bibnamefont {Hartle}}\ and\ \bibinfo {author} {\bibfnamefont {S.~W.}\ \bibnamefont {Hawking}},\ }\href {\doibase 10.1103/PhysRevD.28.2960} {\bibfield  {journal} {\bibinfo  {journal} {Phys. Rev. D}\ }\textbf {\bibinfo {volume} {28}},\ \bibinfo {pages} {2960} (\bibinfo {year} {1983})}\BibitemShut {NoStop}%
\bibitem [{\citenamefont {Vilenkin}(1983)}]{Vilenkin:1983xq}%
  \BibitemOpen
  \bibfield  {author} {\bibinfo {author} {\bibfnamefont {A.}~\bibnamefont {Vilenkin}},\ }\href {\doibase 10.1103/PhysRevD.27.2848} {\bibfield  {journal} {\bibinfo  {journal} {Phys. Rev. D}\ }\textbf {\bibinfo {volume} {27}},\ \bibinfo {pages} {2848} (\bibinfo {year} {1983})}\BibitemShut {NoStop}%
\bibitem [{\citenamefont {Betzios}\ and\ \citenamefont {Papadoulaki}(2024)}]{Betzios:2024oli}%
  \BibitemOpen
  \bibfield  {author} {\bibinfo {author} {\bibfnamefont {P.}~\bibnamefont {Betzios}}\ and\ \bibinfo {author} {\bibfnamefont {O.}~\bibnamefont {Papadoulaki}},\ }\href {\doibase 10.1103/PhysRevLett.133.021501} {\bibfield  {journal} {\bibinfo  {journal} {Phys. Rev. Lett.}\ }\textbf {\bibinfo {volume} {133}},\ \bibinfo {pages} {021501} (\bibinfo {year} {2024})},\ \Eprint {http://arxiv.org/abs/2403.17046} {arXiv:2403.17046 [hep-th]} \BibitemShut {NoStop}%
\bibitem [{\citenamefont {Bezrukov}\ and\ \citenamefont {Shaposhnikov}(2008)}]{Bezrukov:2007ep}%
  \BibitemOpen
  \bibfield  {author} {\bibinfo {author} {\bibfnamefont {F.~L.}\ \bibnamefont {Bezrukov}}\ and\ \bibinfo {author} {\bibfnamefont {M.}~\bibnamefont {Shaposhnikov}},\ }\href {\doibase 10.1016/j.physletb.2007.11.072} {\bibfield  {journal} {\bibinfo  {journal} {Phys. Lett. B}\ }\textbf {\bibinfo {volume} {659}},\ \bibinfo {pages} {703} (\bibinfo {year} {2008})},\ \Eprint {http://arxiv.org/abs/0710.3755} {arXiv:0710.3755 [hep-th]} \BibitemShut {NoStop}%
\bibitem [{\citenamefont {Bezrukov}\ \emph {et~al.}(2012)\citenamefont {Bezrukov}, \citenamefont {Kalmykov}, \citenamefont {Kniehl},\ and\ \citenamefont {Shaposhnikov}}]{Bezrukov:2012sa}%
  \BibitemOpen
  \bibfield  {author} {\bibinfo {author} {\bibfnamefont {F.}~\bibnamefont {Bezrukov}}, \bibinfo {author} {\bibfnamefont {M.~Y.}\ \bibnamefont {Kalmykov}}, \bibinfo {author} {\bibfnamefont {B.~A.}\ \bibnamefont {Kniehl}}, \ and\ \bibinfo {author} {\bibfnamefont {M.}~\bibnamefont {Shaposhnikov}},\ }\href {\doibase 10.1007/JHEP10(2012)140} {\bibfield  {journal} {\bibinfo  {journal} {JHEP}\ }\textbf {\bibinfo {volume} {10}},\ \bibinfo {pages} {140} (\bibinfo {year} {2012})},\ \Eprint {http://arxiv.org/abs/1205.2893} {arXiv:1205.2893 [hep-ph]} \BibitemShut {NoStop}%
\bibitem [{\citenamefont {Buttazzo}\ \emph {et~al.}(2013)\citenamefont {Buttazzo}, \citenamefont {Degrassi}, \citenamefont {Giardino}, \citenamefont {Giudice}, \citenamefont {Sala}, \citenamefont {Salvio},\ and\ \citenamefont {Strumia}}]{Buttazzo:2013uya}%
  \BibitemOpen
  \bibfield  {author} {\bibinfo {author} {\bibfnamefont {D.}~\bibnamefont {Buttazzo}}, \bibinfo {author} {\bibfnamefont {G.}~\bibnamefont {Degrassi}}, \bibinfo {author} {\bibfnamefont {P.~P.}\ \bibnamefont {Giardino}}, \bibinfo {author} {\bibfnamefont {G.~F.}\ \bibnamefont {Giudice}}, \bibinfo {author} {\bibfnamefont {F.}~\bibnamefont {Sala}}, \bibinfo {author} {\bibfnamefont {A.}~\bibnamefont {Salvio}}, \ and\ \bibinfo {author} {\bibfnamefont {A.}~\bibnamefont {Strumia}},\ }\href {\doibase 10.1007/JHEP12(2013)089} {\bibfield  {journal} {\bibinfo  {journal} {JHEP}\ }\textbf {\bibinfo {volume} {12}},\ \bibinfo {pages} {089} (\bibinfo {year} {2013})},\ \Eprint {http://arxiv.org/abs/1307.3536} {arXiv:1307.3536 [hep-ph]} \BibitemShut {NoStop}%
\bibitem [{\citenamefont {Navas}\ \emph {et~al.}(2024)\citenamefont {Navas} \emph {et~al.}}]{ParticleDataGroup:2024cfk}%
  \BibitemOpen
  \bibfield  {author} {\bibinfo {author} {\bibfnamefont {S.}~\bibnamefont {Navas}} \emph {et~al.} (\bibinfo {collaboration} {Particle Data Group}),\ }\href {\doibase 10.1103/PhysRevD.110.030001} {\bibfield  {journal} {\bibinfo  {journal} {Phys. Rev. D}\ }\textbf {\bibinfo {volume} {110}},\ \bibinfo {pages} {030001} (\bibinfo {year} {2024})}\BibitemShut {NoStop}%
\bibitem [{\citenamefont {Starobinsky}(1980)}]{Starobinsky:1980te}%
  \BibitemOpen
  \bibfield  {author} {\bibinfo {author} {\bibfnamefont {A.~A.}\ \bibnamefont {Starobinsky}},\ }\href {\doibase 10.1016/0370-2693(80)90670-X} {\bibfield  {journal} {\bibinfo  {journal} {Phys. Lett. B}\ }\textbf {\bibinfo {volume} {91}},\ \bibinfo {pages} {99} (\bibinfo {year} {1980})}\BibitemShut {NoStop}%
\bibitem [{\citenamefont {Akrami}\ \emph {et~al.}(2020)\citenamefont {Akrami} \emph {et~al.}}]{Planck:2018jri}%
  \BibitemOpen
  \bibfield  {author} {\bibinfo {author} {\bibfnamefont {Y.}~\bibnamefont {Akrami}} \emph {et~al.} (\bibinfo {collaboration} {Planck}),\ }\href {\doibase 10.1051/0004-6361/201833887} {\bibfield  {journal} {\bibinfo  {journal} {Astron. Astrophys.}\ }\textbf {\bibinfo {volume} {641}},\ \bibinfo {pages} {A10} (\bibinfo {year} {2020})},\ \Eprint {http://arxiv.org/abs/1807.06211} {arXiv:1807.06211 [astro-ph.CO]} \BibitemShut {NoStop}%
\bibitem [{\citenamefont {Ade}\ \emph {et~al.}(2021)\citenamefont {Ade} \emph {et~al.}}]{BICEP:2021xfz}%
  \BibitemOpen
  \bibfield  {author} {\bibinfo {author} {\bibfnamefont {P.~A.~R.}\ \bibnamefont {Ade}} \emph {et~al.} (\bibinfo {collaboration} {BICEP, Keck}),\ }\href {\doibase 10.1103/PhysRevLett.127.151301} {\bibfield  {journal} {\bibinfo  {journal} {Phys. Rev. Lett.}\ }\textbf {\bibinfo {volume} {127}},\ \bibinfo {pages} {151301} (\bibinfo {year} {2021})},\ \Eprint {http://arxiv.org/abs/2110.00483} {arXiv:2110.00483 [astro-ph.CO]} \BibitemShut {NoStop}%
\bibitem [{\citenamefont {Barbon}\ and\ \citenamefont {Espinosa}(2009)}]{Barbon:2009ya}%
  \BibitemOpen
  \bibfield  {author} {\bibinfo {author} {\bibfnamefont {J.~L.~F.}\ \bibnamefont {Barbon}}\ and\ \bibinfo {author} {\bibfnamefont {J.~R.}\ \bibnamefont {Espinosa}},\ }\href {\doibase 10.1103/PhysRevD.79.081302} {\bibfield  {journal} {\bibinfo  {journal} {Phys. Rev. D}\ }\textbf {\bibinfo {volume} {79}},\ \bibinfo {pages} {081302} (\bibinfo {year} {2009})},\ \Eprint {http://arxiv.org/abs/0903.0355} {arXiv:0903.0355 [hep-ph]} \BibitemShut {NoStop}%
\bibitem [{\citenamefont {Bezrukov}\ \emph {et~al.}(2011)\citenamefont {Bezrukov}, \citenamefont {Magnin}, \citenamefont {Shaposhnikov},\ and\ \citenamefont {Sibiryakov}}]{Bezrukov:2010jz}%
  \BibitemOpen
  \bibfield  {author} {\bibinfo {author} {\bibfnamefont {F.}~\bibnamefont {Bezrukov}}, \bibinfo {author} {\bibfnamefont {A.}~\bibnamefont {Magnin}}, \bibinfo {author} {\bibfnamefont {M.}~\bibnamefont {Shaposhnikov}}, \ and\ \bibinfo {author} {\bibfnamefont {S.}~\bibnamefont {Sibiryakov}},\ }\href {\doibase 10.1007/JHEP01(2011)016} {\bibfield  {journal} {\bibinfo  {journal} {JHEP}\ }\textbf {\bibinfo {volume} {01}},\ \bibinfo {pages} {016} (\bibinfo {year} {2011})},\ \Eprint {http://arxiv.org/abs/1008.5157} {arXiv:1008.5157 [hep-ph]} \BibitemShut {NoStop}%
\bibitem [{\citenamefont {Turner}\ and\ \citenamefont {Widrow}(1988)}]{Turner:1987bw}%
  \BibitemOpen
  \bibfield  {author} {\bibinfo {author} {\bibfnamefont {M.~S.}\ \bibnamefont {Turner}}\ and\ \bibinfo {author} {\bibfnamefont {L.~M.}\ \bibnamefont {Widrow}},\ }\href {\doibase 10.1103/PhysRevD.37.2743} {\bibfield  {journal} {\bibinfo  {journal} {Phys. Rev. D}\ }\textbf {\bibinfo {volume} {37}},\ \bibinfo {pages} {2743} (\bibinfo {year} {1988})}\BibitemShut {NoStop}%
\bibitem [{\citenamefont {Hogan}(1983)}]{Hogan:1983zz}%
  \BibitemOpen
  \bibfield  {author} {\bibinfo {author} {\bibfnamefont {C.~J.}\ \bibnamefont {Hogan}},\ }\href {\doibase 10.1103/PhysRevLett.51.1488} {\bibfield  {journal} {\bibinfo  {journal} {Phys. Rev. Lett.}\ }\textbf {\bibinfo {volume} {51}},\ \bibinfo {pages} {1488} (\bibinfo {year} {1983})}\BibitemShut {NoStop}%
\bibitem [{\citenamefont {Kandus}\ \emph {et~al.}(2011)\citenamefont {Kandus}, \citenamefont {Kunze},\ and\ \citenamefont {Tsagas}}]{Kandus:2010nw}%
  \BibitemOpen
  \bibfield  {author} {\bibinfo {author} {\bibfnamefont {A.}~\bibnamefont {Kandus}}, \bibinfo {author} {\bibfnamefont {K.~E.}\ \bibnamefont {Kunze}}, \ and\ \bibinfo {author} {\bibfnamefont {C.~G.}\ \bibnamefont {Tsagas}},\ }\href {\doibase 10.1016/j.physrep.2011.03.001} {\bibfield  {journal} {\bibinfo  {journal} {Phys. Rept.}\ }\textbf {\bibinfo {volume} {505}},\ \bibinfo {pages} {1} (\bibinfo {year} {2011})},\ \Eprint {http://arxiv.org/abs/1007.3891} {arXiv:1007.3891 [astro-ph.CO]} \BibitemShut {NoStop}%
\bibitem [{\citenamefont {Durrer}\ and\ \citenamefont {Neronov}(2013)}]{Durrer:2013pga}%
  \BibitemOpen
  \bibfield  {author} {\bibinfo {author} {\bibfnamefont {R.}~\bibnamefont {Durrer}}\ and\ \bibinfo {author} {\bibfnamefont {A.}~\bibnamefont {Neronov}},\ }\href {\doibase 10.1007/s00159-013-0062-7} {\bibfield  {journal} {\bibinfo  {journal} {Astron. Astrophys. Rev.}\ }\textbf {\bibinfo {volume} {21}},\ \bibinfo {pages} {62} (\bibinfo {year} {2013})},\ \Eprint {http://arxiv.org/abs/1303.7121} {arXiv:1303.7121 [astro-ph.CO]} \BibitemShut {NoStop}%
\bibitem [{\citenamefont {Subramanian}(2016)}]{Subramanian:2015lua}%
  \BibitemOpen
  \bibfield  {author} {\bibinfo {author} {\bibfnamefont {K.}~\bibnamefont {Subramanian}},\ }\href {\doibase 10.1088/0034-4885/79/7/076901} {\bibfield  {journal} {\bibinfo  {journal} {Rept. Prog. Phys.}\ }\textbf {\bibinfo {volume} {79}},\ \bibinfo {pages} {076901} (\bibinfo {year} {2016})},\ \Eprint {http://arxiv.org/abs/1504.02311} {arXiv:1504.02311 [astro-ph.CO]} \BibitemShut {NoStop}%
\bibitem [{\citenamefont {Maldacena}\ and\ \citenamefont {Maoz}(2004)}]{Maldacena:2004rf}%
  \BibitemOpen
  \bibfield  {author} {\bibinfo {author} {\bibfnamefont {J.~M.}\ \bibnamefont {Maldacena}}\ and\ \bibinfo {author} {\bibfnamefont {L.}~\bibnamefont {Maoz}},\ }\href {\doibase 10.1088/1126-6708/2004/02/053} {\bibfield  {journal} {\bibinfo  {journal} {JHEP}\ }\textbf {\bibinfo {volume} {02}},\ \bibinfo {pages} {053} (\bibinfo {year} {2004})},\ \Eprint {http://arxiv.org/abs/hep-th/0401024} {arXiv:hep-th/0401024} \BibitemShut {NoStop}%
\bibitem [{\citenamefont {Betzios}\ \emph {et~al.}(2019)\citenamefont {Betzios}, \citenamefont {Kiritsis},\ and\ \citenamefont {Papadoulaki}}]{Betzios:2019rds}%
  \BibitemOpen
  \bibfield  {author} {\bibinfo {author} {\bibfnamefont {P.}~\bibnamefont {Betzios}}, \bibinfo {author} {\bibfnamefont {E.}~\bibnamefont {Kiritsis}}, \ and\ \bibinfo {author} {\bibfnamefont {O.}~\bibnamefont {Papadoulaki}},\ }\href {\doibase 10.1007/JHEP06(2019)042} {\bibfield  {journal} {\bibinfo  {journal} {JHEP}\ }\textbf {\bibinfo {volume} {06}},\ \bibinfo {pages} {042} (\bibinfo {year} {2019})},\ \Eprint {http://arxiv.org/abs/1903.05658} {arXiv:1903.05658 [hep-th]} \BibitemShut {NoStop}%
\bibitem [{\citenamefont {Betzios}\ \emph {et~al.}(2022)\citenamefont {Betzios}, \citenamefont {Kiritsis},\ and\ \citenamefont {Papadoulaki}}]{Betzios:2021fnm}%
  \BibitemOpen
  \bibfield  {author} {\bibinfo {author} {\bibfnamefont {P.}~\bibnamefont {Betzios}}, \bibinfo {author} {\bibfnamefont {E.}~\bibnamefont {Kiritsis}}, \ and\ \bibinfo {author} {\bibfnamefont {O.}~\bibnamefont {Papadoulaki}},\ }\href {\doibase 10.1007/JHEP02(2022)126} {\bibfield  {journal} {\bibinfo  {journal} {JHEP}\ }\textbf {\bibinfo {volume} {02}},\ \bibinfo {pages} {126} (\bibinfo {year} {2022})},\ \Eprint {http://arxiv.org/abs/2110.14655} {arXiv:2110.14655 [hep-th]} \BibitemShut {NoStop}%
\bibitem [{\citenamefont {Van~Raamsdonk}(2022)}]{VanRaamsdonk:2021qgv}%
  \BibitemOpen
  \bibfield  {author} {\bibinfo {author} {\bibfnamefont {M.}~\bibnamefont {Van~Raamsdonk}},\ }\href {\doibase 10.1007/JHEP03(2022)039} {\bibfield  {journal} {\bibinfo  {journal} {JHEP}\ }\textbf {\bibinfo {volume} {03}},\ \bibinfo {pages} {039} (\bibinfo {year} {2022})},\ \Eprint {http://arxiv.org/abs/2102.05057} {arXiv:2102.05057 [hep-th]} \BibitemShut {NoStop}%
\bibitem [{\citenamefont {Antonini}\ \emph {et~al.}(2023)\citenamefont {Antonini}, \citenamefont {Simidzija}, \citenamefont {Swingle},\ and\ \citenamefont {Van~Raamsdonk}}]{Antonini:2022ptt}%
  \BibitemOpen
  \bibfield  {author} {\bibinfo {author} {\bibfnamefont {S.}~\bibnamefont {Antonini}}, \bibinfo {author} {\bibfnamefont {P.}~\bibnamefont {Simidzija}}, \bibinfo {author} {\bibfnamefont {B.}~\bibnamefont {Swingle}}, \ and\ \bibinfo {author} {\bibfnamefont {M.}~\bibnamefont {Van~Raamsdonk}},\ }\href {\doibase 10.1103/PhysRevLett.130.221601} {\bibfield  {journal} {\bibinfo  {journal} {Phys. Rev. Lett.}\ }\textbf {\bibinfo {volume} {130}},\ \bibinfo {pages} {221601} (\bibinfo {year} {2023})},\ \Eprint {http://arxiv.org/abs/2206.14821} {arXiv:2206.14821 [hep-th]} \BibitemShut {NoStop}%
\bibitem [{\citenamefont {Birrell}\ and\ \citenamefont {Davies}(1982)}]{Birrell:1982ix}%
  \BibitemOpen
  \bibfield  {author} {\bibinfo {author} {\bibfnamefont {N.~D.}\ \bibnamefont {Birrell}}\ and\ \bibinfo {author} {\bibfnamefont {P.~C.~W.}\ \bibnamefont {Davies}},\ }\href {\doibase 10.1017/CBO9780511622632} {\emph {\bibinfo {title} {{Quantum Fields in Curved Space}}}},\ Cambridge Monographs on Mathematical Physics\ (\bibinfo  {publisher} {Cambridge University Press},\ \bibinfo {address} {Cambridge, UK},\ \bibinfo {year} {1982})\BibitemShut {NoStop}%
\bibitem [{\citenamefont {Rey}(1990)}]{Rey:1989th}%
  \BibitemOpen
  \bibfield  {author} {\bibinfo {author} {\bibfnamefont {S.-J.}\ \bibnamefont {Rey}},\ }\href {\doibase 10.1016/0550-3213(90)90346-F} {\bibfield  {journal} {\bibinfo  {journal} {Nucl. Phys. B}\ }\textbf {\bibinfo {volume} {336}},\ \bibinfo {pages} {146} (\bibinfo {year} {1990})}\BibitemShut {NoStop}%
\bibitem [{\citenamefont {Hosoya}\ and\ \citenamefont {Ogura}(1989)}]{Hosoya:1989zn}%
  \BibitemOpen
  \bibfield  {author} {\bibinfo {author} {\bibfnamefont {A.}~\bibnamefont {Hosoya}}\ and\ \bibinfo {author} {\bibfnamefont {W.}~\bibnamefont {Ogura}},\ }\href {\doibase 10.1016/0370-2693(89)91020-4} {\bibfield  {journal} {\bibinfo  {journal} {Phys. Lett. B}\ }\textbf {\bibinfo {volume} {225}},\ \bibinfo {pages} {117} (\bibinfo {year} {1989})}\BibitemShut {NoStop}%
\bibitem [{\citenamefont {Marolf}\ and\ \citenamefont {Santos}(2021)}]{Marolf:2021kjc}%
  \BibitemOpen
  \bibfield  {author} {\bibinfo {author} {\bibfnamefont {D.}~\bibnamefont {Marolf}}\ and\ \bibinfo {author} {\bibfnamefont {J.~E.}\ \bibnamefont {Santos}},\ }\href {\doibase 10.1088/1361-6382/ac2cb7} {\bibfield  {journal} {\bibinfo  {journal} {Class. Quant. Grav.}\ }\textbf {\bibinfo {volume} {38}},\ \bibinfo {pages} {224002} (\bibinfo {year} {2021})},\ \Eprint {http://arxiv.org/abs/2101.08875} {arXiv:2101.08875 [hep-th]} \BibitemShut {NoStop}%
\bibitem [{\citenamefont {Deshpande}\ and\ \citenamefont {Lunin}(2023)}]{Deshpande:2022zfm}%
  \BibitemOpen
  \bibfield  {author} {\bibinfo {author} {\bibfnamefont {R.}~\bibnamefont {Deshpande}}\ and\ \bibinfo {author} {\bibfnamefont {O.}~\bibnamefont {Lunin}},\ }\href {\doibase 10.1016/j.nuclphysb.2023.116355} {\bibfield  {journal} {\bibinfo  {journal} {Nucl. Phys. B}\ }\textbf {\bibinfo {volume} {996}},\ \bibinfo {pages} {116355} (\bibinfo {year} {2023})},\ \Eprint {http://arxiv.org/abs/2212.11962} {arXiv:2212.11962 [hep-th]} \BibitemShut {NoStop}%
\bibitem [{\citenamefont {Lan}\ and\ \citenamefont {Piao}(2024)}]{Lan:2024gnv}%
  \BibitemOpen
  \bibfield  {author} {\bibinfo {author} {\bibfnamefont {Q.-Y.}\ \bibnamefont {Lan}}\ and\ \bibinfo {author} {\bibfnamefont {Y.-S.}\ \bibnamefont {Piao}},\ }\href@noop {} {\  (\bibinfo {year} {2024})},\ \Eprint {http://arxiv.org/abs/2411.13844} {arXiv:2411.13844 [gr-qc]} \BibitemShut {NoStop}%
\bibitem [{\citenamefont {Gupta}\ \emph {et~al.}(1990)\citenamefont {Gupta}, \citenamefont {Hughes}, \citenamefont {Preskill},\ and\ \citenamefont {Wise}}]{Gupta:1989bs}%
  \BibitemOpen
  \bibfield  {author} {\bibinfo {author} {\bibfnamefont {A.~K.}\ \bibnamefont {Gupta}}, \bibinfo {author} {\bibfnamefont {J.}~\bibnamefont {Hughes}}, \bibinfo {author} {\bibfnamefont {J.}~\bibnamefont {Preskill}}, \ and\ \bibinfo {author} {\bibfnamefont {M.~B.}\ \bibnamefont {Wise}},\ }\href {\doibase 10.1016/0550-3213(90)90228-6} {\bibfield  {journal} {\bibinfo  {journal} {Nucl. Phys. B}\ }\textbf {\bibinfo {volume} {333}},\ \bibinfo {pages} {195} (\bibinfo {year} {1990})}\BibitemShut {NoStop}%
\bibitem [{\citenamefont {Coleman}(1985)}]{Coleman:1985rnk}%
  \BibitemOpen
  \bibfield  {author} {\bibinfo {author} {\bibfnamefont {S.}~\bibnamefont {Coleman}},\ }\href {\doibase 10.1017/CBO9780511565045} {\emph {\bibinfo {title} {{Aspects of Symmetry}: {Selected Erice Lectures}}}}\ (\bibinfo  {publisher} {Cambridge University Press},\ \bibinfo {address} {Cambridge, U.K.},\ \bibinfo {year} {1985})\BibitemShut {NoStop}%
\bibitem [{\citenamefont {Giovannini}(2004)}]{Giovannini:2003yn}%
  \BibitemOpen
  \bibfield  {author} {\bibinfo {author} {\bibfnamefont {M.}~\bibnamefont {Giovannini}},\ }\href {\doibase 10.1142/S0218271804004530} {\bibfield  {journal} {\bibinfo  {journal} {Int. J. Mod. Phys. D}\ }\textbf {\bibinfo {volume} {13}},\ \bibinfo {pages} {391} (\bibinfo {year} {2004})},\ \Eprint {http://arxiv.org/abs/astro-ph/0312614} {arXiv:astro-ph/0312614} \BibitemShut {NoStop}%
\bibitem [{\citenamefont {Brandenburg}\ and\ \citenamefont {Subramanian}(2005)}]{Brandenburg:2004jv}%
  \BibitemOpen
  \bibfield  {author} {\bibinfo {author} {\bibfnamefont {A.}~\bibnamefont {Brandenburg}}\ and\ \bibinfo {author} {\bibfnamefont {K.}~\bibnamefont {Subramanian}},\ }\href {\doibase 10.1016/j.physrep.2005.06.005} {\bibfield  {journal} {\bibinfo  {journal} {Phys. Rept.}\ }\textbf {\bibinfo {volume} {417}},\ \bibinfo {pages} {1} (\bibinfo {year} {2005})},\ \Eprint {http://arxiv.org/abs/astro-ph/0405052} {arXiv:astro-ph/0405052} \BibitemShut {NoStop}%
\bibitem [{\citenamefont {Barrow}\ \emph {et~al.}(2007)\citenamefont {Barrow}, \citenamefont {Maartens},\ and\ \citenamefont {Tsagas}}]{Barrow:2006ch}%
  \BibitemOpen
  \bibfield  {author} {\bibinfo {author} {\bibfnamefont {J.~D.}\ \bibnamefont {Barrow}}, \bibinfo {author} {\bibfnamefont {R.}~\bibnamefont {Maartens}}, \ and\ \bibinfo {author} {\bibfnamefont {C.~G.}\ \bibnamefont {Tsagas}},\ }\href {\doibase 10.1016/j.physrep.2007.04.006} {\bibfield  {journal} {\bibinfo  {journal} {Phys. Rept.}\ }\textbf {\bibinfo {volume} {449}},\ \bibinfo {pages} {131} (\bibinfo {year} {2007})},\ \Eprint {http://arxiv.org/abs/astro-ph/0611537} {arXiv:astro-ph/0611537} \BibitemShut {NoStop}%
\bibitem [{\citenamefont {Masina}\ and\ \citenamefont {Quiros}(2024)}]{Masina:2024ybn}%
  \BibitemOpen
  \bibfield  {author} {\bibinfo {author} {\bibfnamefont {I.}~\bibnamefont {Masina}}\ and\ \bibinfo {author} {\bibfnamefont {M.}~\bibnamefont {Quiros}},\ }\href {\doibase 10.1140/epjc/s10052-024-13522-x} {\bibfield  {journal} {\bibinfo  {journal} {Eur. Phys. J. C}\ }\textbf {\bibinfo {volume} {84}},\ \bibinfo {pages} {1153} (\bibinfo {year} {2024})},\ \Eprint {http://arxiv.org/abs/2403.02461} {arXiv:2403.02461 [hep-ph]} \BibitemShut {NoStop}%
\bibitem [{\citenamefont {Bezrukov}\ \emph {et~al.}(2015)\citenamefont {Bezrukov}, \citenamefont {Rubio},\ and\ \citenamefont {Shaposhnikov}}]{Bezrukov:2014ipa}%
  \BibitemOpen
  \bibfield  {author} {\bibinfo {author} {\bibfnamefont {F.}~\bibnamefont {Bezrukov}}, \bibinfo {author} {\bibfnamefont {J.}~\bibnamefont {Rubio}}, \ and\ \bibinfo {author} {\bibfnamefont {M.}~\bibnamefont {Shaposhnikov}},\ }\href {\doibase 10.1103/PhysRevD.92.083512} {\bibfield  {journal} {\bibinfo  {journal} {Phys. Rev. D}\ }\textbf {\bibinfo {volume} {92}},\ \bibinfo {pages} {083512} (\bibinfo {year} {2015})},\ \Eprint {http://arxiv.org/abs/1412.3811} {arXiv:1412.3811 [hep-ph]} \BibitemShut {NoStop}%
\bibitem [{\citenamefont {Fumagalli}\ and\ \citenamefont {Postma}(2016)}]{Fumagalli:2016lls}%
  \BibitemOpen
  \bibfield  {author} {\bibinfo {author} {\bibfnamefont {J.}~\bibnamefont {Fumagalli}}\ and\ \bibinfo {author} {\bibfnamefont {M.}~\bibnamefont {Postma}},\ }\href {\doibase 10.1007/JHEP05(2016)049} {\bibfield  {journal} {\bibinfo  {journal} {JHEP}\ }\textbf {\bibinfo {volume} {05}},\ \bibinfo {pages} {049} (\bibinfo {year} {2016})},\ \Eprint {http://arxiv.org/abs/1602.07234} {arXiv:1602.07234 [hep-ph]} \BibitemShut {NoStop}%
\bibitem [{\citenamefont {Enckell}\ \emph {et~al.}(2018)\citenamefont {Enckell}, \citenamefont {Enqvist}, \citenamefont {Rasanen},\ and\ \citenamefont {Tomberg}}]{Enckell:2018kkc}%
  \BibitemOpen
  \bibfield  {author} {\bibinfo {author} {\bibfnamefont {V.-M.}\ \bibnamefont {Enckell}}, \bibinfo {author} {\bibfnamefont {K.}~\bibnamefont {Enqvist}}, \bibinfo {author} {\bibfnamefont {S.}~\bibnamefont {Rasanen}}, \ and\ \bibinfo {author} {\bibfnamefont {E.}~\bibnamefont {Tomberg}},\ }\href {\doibase 10.1088/1475-7516/2018/06/005} {\bibfield  {journal} {\bibinfo  {journal} {JCAP}\ }\textbf {\bibinfo {volume} {06}},\ \bibinfo {pages} {005} (\bibinfo {year} {2018})},\ \Eprint {http://arxiv.org/abs/1802.09299} {arXiv:1802.09299 [astro-ph.CO]} \BibitemShut {NoStop}%
\bibitem [{\citenamefont {Baumann}\ and\ \citenamefont {McAllister}(2015)}]{Baumann:2014nda}%
  \BibitemOpen
  \bibfield  {author} {\bibinfo {author} {\bibfnamefont {D.}~\bibnamefont {Baumann}}\ and\ \bibinfo {author} {\bibfnamefont {L.}~\bibnamefont {McAllister}},\ }\href {\doibase 10.1017/CBO9781316105733} {\emph {\bibinfo {title} {{Inflation and String Theory}}}},\ Cambridge Monographs on Mathematical Physics\ (\bibinfo  {publisher} {Cambridge University Press},\ \bibinfo {year} {2015})\ \Eprint {http://arxiv.org/abs/1404.2601} {arXiv:1404.2601 [hep-th]} \BibitemShut {NoStop}%
\bibitem [{\citenamefont {Rubio}(2019)}]{Rubio:2018ogq}%
  \BibitemOpen
  \bibfield  {author} {\bibinfo {author} {\bibfnamefont {J.}~\bibnamefont {Rubio}},\ }\href {\doibase 10.3389/fspas.2018.00050} {\bibfield  {journal} {\bibinfo  {journal} {Front. Astron. Space Sci.}\ }\textbf {\bibinfo {volume} {5}},\ \bibinfo {pages} {50} (\bibinfo {year} {2019})},\ \Eprint {http://arxiv.org/abs/1807.02376} {arXiv:1807.02376 [hep-ph]} \BibitemShut {NoStop}%
\bibitem [{\citenamefont {Burgess}\ \emph {et~al.}(2009)\citenamefont {Burgess}, \citenamefont {Lee},\ and\ \citenamefont {Trott}}]{Burgess:2009ea}%
  \BibitemOpen
  \bibfield  {author} {\bibinfo {author} {\bibfnamefont {C.~P.}\ \bibnamefont {Burgess}}, \bibinfo {author} {\bibfnamefont {H.~M.}\ \bibnamefont {Lee}}, \ and\ \bibinfo {author} {\bibfnamefont {M.}~\bibnamefont {Trott}},\ }\href {\doibase 10.1088/1126-6708/2009/09/103} {\bibfield  {journal} {\bibinfo  {journal} {JHEP}\ }\textbf {\bibinfo {volume} {09}},\ \bibinfo {pages} {103} (\bibinfo {year} {2009})},\ \Eprint {http://arxiv.org/abs/0902.4465} {arXiv:0902.4465 [hep-ph]} \BibitemShut {NoStop}%
\bibitem [{\citenamefont {Burgess}\ \emph {et~al.}(2010)\citenamefont {Burgess}, \citenamefont {Lee},\ and\ \citenamefont {Trott}}]{Burgess:2010zq}%
  \BibitemOpen
  \bibfield  {author} {\bibinfo {author} {\bibfnamefont {C.~P.}\ \bibnamefont {Burgess}}, \bibinfo {author} {\bibfnamefont {H.~M.}\ \bibnamefont {Lee}}, \ and\ \bibinfo {author} {\bibfnamefont {M.}~\bibnamefont {Trott}},\ }\href {\doibase 10.1007/JHEP07(2010)007} {\bibfield  {journal} {\bibinfo  {journal} {JHEP}\ }\textbf {\bibinfo {volume} {07}},\ \bibinfo {pages} {007} (\bibinfo {year} {2010})},\ \Eprint {http://arxiv.org/abs/1002.2730} {arXiv:1002.2730 [hep-ph]} \BibitemShut {NoStop}%
\bibitem [{\citenamefont {Kulsrud}\ and\ \citenamefont {Anderson}(1992)}]{Kulsrud:1992rk}%
  \BibitemOpen
  \bibfield  {author} {\bibinfo {author} {\bibfnamefont {R.~M.}\ \bibnamefont {Kulsrud}}\ and\ \bibinfo {author} {\bibfnamefont {S.~W.}\ \bibnamefont {Anderson}},\ }\href {\doibase 10.1086/171743} {\bibfield  {journal} {\bibinfo  {journal} {Astrophys. J.}\ }\textbf {\bibinfo {volume} {396}},\ \bibinfo {pages} {606} (\bibinfo {year} {1992})}\BibitemShut {NoStop}%
\bibitem [{\citenamefont {Beck}\ \emph {et~al.}(1996)\citenamefont {Beck}, \citenamefont {Brandenburg}, \citenamefont {Moss}, \citenamefont {Shukurov},\ and\ \citenamefont {Sokoloff}}]{Beck:1995zs}%
  \BibitemOpen
  \bibfield  {author} {\bibinfo {author} {\bibfnamefont {R.}~\bibnamefont {Beck}}, \bibinfo {author} {\bibfnamefont {A.}~\bibnamefont {Brandenburg}}, \bibinfo {author} {\bibfnamefont {D.}~\bibnamefont {Moss}}, \bibinfo {author} {\bibfnamefont {A.}~\bibnamefont {Shukurov}}, \ and\ \bibinfo {author} {\bibfnamefont {D.}~\bibnamefont {Sokoloff}},\ }\href {\doibase 10.1146/annurev.astro.34.1.155} {\bibfield  {journal} {\bibinfo  {journal} {Ann. Rev. Astron. Astrophys.}\ }\textbf {\bibinfo {volume} {34}},\ \bibinfo {pages} {155} (\bibinfo {year} {1996})}\BibitemShut {NoStop}%
\bibitem [{\citenamefont {Kulsrud}(1999)}]{Kulsrud:1999bg}%
  \BibitemOpen
  \bibfield  {author} {\bibinfo {author} {\bibfnamefont {R.~M.}\ \bibnamefont {Kulsrud}},\ }\href {\doibase 10.1146/annurev.astro.37.1.37} {\bibfield  {journal} {\bibinfo  {journal} {Ann. Rev. Astron. Astrophys.}\ }\textbf {\bibinfo {volume} {37}},\ \bibinfo {pages} {37} (\bibinfo {year} {1999})}\BibitemShut {NoStop}%
\bibitem [{\citenamefont {Widrow}(2002)}]{Widrow:2002ud}%
  \BibitemOpen
  \bibfield  {author} {\bibinfo {author} {\bibfnamefont {L.~M.}\ \bibnamefont {Widrow}},\ }\href {\doibase 10.1103/RevModPhys.74.775} {\bibfield  {journal} {\bibinfo  {journal} {Rev. Mod. Phys.}\ }\textbf {\bibinfo {volume} {74}},\ \bibinfo {pages} {775} (\bibinfo {year} {2002})},\ \Eprint {http://arxiv.org/abs/astro-ph/0207240} {arXiv:astro-ph/0207240} \BibitemShut {NoStop}%
\bibitem [{\citenamefont {Almheiri}\ \emph {et~al.}(2021)\citenamefont {Almheiri}, \citenamefont {Hartman}, \citenamefont {Maldacena}, \citenamefont {Shaghoulian},\ and\ \citenamefont {Tajdini}}]{Almheiri:2020cfm}%
  \BibitemOpen
  \bibfield  {author} {\bibinfo {author} {\bibfnamefont {A.}~\bibnamefont {Almheiri}}, \bibinfo {author} {\bibfnamefont {T.}~\bibnamefont {Hartman}}, \bibinfo {author} {\bibfnamefont {J.}~\bibnamefont {Maldacena}}, \bibinfo {author} {\bibfnamefont {E.}~\bibnamefont {Shaghoulian}}, \ and\ \bibinfo {author} {\bibfnamefont {A.}~\bibnamefont {Tajdini}},\ }\href {\doibase 10.1103/RevModPhys.93.035002} {\bibfield  {journal} {\bibinfo  {journal} {Rev. Mod. Phys.}\ }\textbf {\bibinfo {volume} {93}},\ \bibinfo {pages} {035002} (\bibinfo {year} {2021})},\ \Eprint {http://arxiv.org/abs/2006.06872} {arXiv:2006.06872 [hep-th]} \BibitemShut {NoStop}%
\bibitem [{\citenamefont {Masina}(2013)}]{Masina:2012tz}%
  \BibitemOpen
  \bibfield  {author} {\bibinfo {author} {\bibfnamefont {I.}~\bibnamefont {Masina}},\ }\href {\doibase 10.1103/PhysRevD.87.053001} {\bibfield  {journal} {\bibinfo  {journal} {Phys. Rev. D}\ }\textbf {\bibinfo {volume} {87}},\ \bibinfo {pages} {053001} (\bibinfo {year} {2013})},\ \Eprint {http://arxiv.org/abs/1209.0393} {arXiv:1209.0393 [hep-ph]} \BibitemShut {NoStop}%
\bibitem [{\citenamefont {Espinosa}\ \emph {et~al.}(2015)\citenamefont {Espinosa}, \citenamefont {Giudice}, \citenamefont {Morgante}, \citenamefont {Riotto}, \citenamefont {Senatore}, \citenamefont {Strumia},\ and\ \citenamefont {Tetradis}}]{Espinosa:2015qea}%
  \BibitemOpen
  \bibfield  {author} {\bibinfo {author} {\bibfnamefont {J.~R.}\ \bibnamefont {Espinosa}}, \bibinfo {author} {\bibfnamefont {G.~F.}\ \bibnamefont {Giudice}}, \bibinfo {author} {\bibfnamefont {E.}~\bibnamefont {Morgante}}, \bibinfo {author} {\bibfnamefont {A.}~\bibnamefont {Riotto}}, \bibinfo {author} {\bibfnamefont {L.}~\bibnamefont {Senatore}}, \bibinfo {author} {\bibfnamefont {A.}~\bibnamefont {Strumia}}, \ and\ \bibinfo {author} {\bibfnamefont {N.}~\bibnamefont {Tetradis}},\ }\href {\doibase 10.1007/JHEP09(2015)174} {\bibfield  {journal} {\bibinfo  {journal} {JHEP}\ }\textbf {\bibinfo {volume} {09}},\ \bibinfo {pages} {174} (\bibinfo {year} {2015})},\ \Eprint {http://arxiv.org/abs/1505.04825} {arXiv:1505.04825 [hep-ph]} \BibitemShut {NoStop}%
\bibitem [{\citenamefont {Bednyakov}\ \emph {et~al.}(2015)\citenamefont {Bednyakov}, \citenamefont {Kniehl}, \citenamefont {Pikelner},\ and\ \citenamefont {Veretin}}]{Bednyakov:2015sca}%
  \BibitemOpen
  \bibfield  {author} {\bibinfo {author} {\bibfnamefont {A.~V.}\ \bibnamefont {Bednyakov}}, \bibinfo {author} {\bibfnamefont {B.~A.}\ \bibnamefont {Kniehl}}, \bibinfo {author} {\bibfnamefont {A.~F.}\ \bibnamefont {Pikelner}}, \ and\ \bibinfo {author} {\bibfnamefont {O.~L.}\ \bibnamefont {Veretin}},\ }\href {\doibase 10.1103/PhysRevLett.115.201802} {\bibfield  {journal} {\bibinfo  {journal} {Phys. Rev. Lett.}\ }\textbf {\bibinfo {volume} {115}},\ \bibinfo {pages} {201802} (\bibinfo {year} {2015})},\ \Eprint {http://arxiv.org/abs/1507.08833} {arXiv:1507.08833 [hep-ph]} \BibitemShut {NoStop}%
\bibitem [{\citenamefont {Coleman}\ and\ \citenamefont {De~Luccia}(1980)}]{Coleman:1980aw}%
  \BibitemOpen
  \bibfield  {author} {\bibinfo {author} {\bibfnamefont {S.~R.}\ \bibnamefont {Coleman}}\ and\ \bibinfo {author} {\bibfnamefont {F.}~\bibnamefont {De~Luccia}},\ }\href {\doibase 10.1103/PhysRevD.21.3305} {\bibfield  {journal} {\bibinfo  {journal} {Phys. Rev. D}\ }\textbf {\bibinfo {volume} {21}},\ \bibinfo {pages} {3305} (\bibinfo {year} {1980})}\BibitemShut {NoStop}%
\bibitem [{\citenamefont {Isidori}\ \emph {et~al.}(2008)\citenamefont {Isidori}, \citenamefont {Rychkov}, \citenamefont {Strumia},\ and\ \citenamefont {Tetradis}}]{Isidori:2007vm}%
  \BibitemOpen
  \bibfield  {author} {\bibinfo {author} {\bibfnamefont {G.}~\bibnamefont {Isidori}}, \bibinfo {author} {\bibfnamefont {V.~S.}\ \bibnamefont {Rychkov}}, \bibinfo {author} {\bibfnamefont {A.}~\bibnamefont {Strumia}}, \ and\ \bibinfo {author} {\bibfnamefont {N.}~\bibnamefont {Tetradis}},\ }\href {\doibase 10.1103/PhysRevD.77.025034} {\bibfield  {journal} {\bibinfo  {journal} {Phys. Rev. D}\ }\textbf {\bibinfo {volume} {77}},\ \bibinfo {pages} {025034} (\bibinfo {year} {2008})},\ \Eprint {http://arxiv.org/abs/0712.0242} {arXiv:0712.0242 [hep-ph]} \BibitemShut {NoStop}%
\bibitem [{\citenamefont {Salvio}\ \emph {et~al.}(2016)\citenamefont {Salvio}, \citenamefont {Strumia}, \citenamefont {Tetradis},\ and\ \citenamefont {Urbano}}]{Salvio:2016mvj}%
  \BibitemOpen
  \bibfield  {author} {\bibinfo {author} {\bibfnamefont {A.}~\bibnamefont {Salvio}}, \bibinfo {author} {\bibfnamefont {A.}~\bibnamefont {Strumia}}, \bibinfo {author} {\bibfnamefont {N.}~\bibnamefont {Tetradis}}, \ and\ \bibinfo {author} {\bibfnamefont {A.}~\bibnamefont {Urbano}},\ }\href {\doibase 10.1007/JHEP09(2016)054} {\bibfield  {journal} {\bibinfo  {journal} {JHEP}\ }\textbf {\bibinfo {volume} {09}},\ \bibinfo {pages} {054} (\bibinfo {year} {2016})},\ \Eprint {http://arxiv.org/abs/1608.02555} {arXiv:1608.02555 [hep-ph]} \BibitemShut {NoStop}%
\bibitem [{\citenamefont {Branchina}\ \emph {et~al.}(2016)\citenamefont {Branchina}, \citenamefont {Messina},\ and\ \citenamefont {Zappala}}]{Branchina:2016bws}%
  \BibitemOpen
  \bibfield  {author} {\bibinfo {author} {\bibfnamefont {V.}~\bibnamefont {Branchina}}, \bibinfo {author} {\bibfnamefont {E.}~\bibnamefont {Messina}}, \ and\ \bibinfo {author} {\bibfnamefont {D.}~\bibnamefont {Zappala}},\ }\href {\doibase 10.1209/0295-5075/116/21001} {\bibfield  {journal} {\bibinfo  {journal} {EPL}\ }\textbf {\bibinfo {volume} {116}},\ \bibinfo {pages} {21001} (\bibinfo {year} {2016})},\ \Eprint {http://arxiv.org/abs/1601.06963} {arXiv:1601.06963 [hep-ph]} \BibitemShut {NoStop}%
\bibitem [{\citenamefont {Markkanen}\ \emph {et~al.}(2018{\natexlab{a}})\citenamefont {Markkanen}, \citenamefont {Rajantie},\ and\ \citenamefont {Stopyra}}]{Markkanen:2018pdo}%
  \BibitemOpen
  \bibfield  {author} {\bibinfo {author} {\bibfnamefont {T.}~\bibnamefont {Markkanen}}, \bibinfo {author} {\bibfnamefont {A.}~\bibnamefont {Rajantie}}, \ and\ \bibinfo {author} {\bibfnamefont {S.}~\bibnamefont {Stopyra}},\ }\href {\doibase 10.3389/fspas.2018.00040} {\bibfield  {journal} {\bibinfo  {journal} {Front. Astron. Space Sci.}\ }\textbf {\bibinfo {volume} {5}},\ \bibinfo {pages} {40} (\bibinfo {year} {2018}{\natexlab{a}})},\ \Eprint {http://arxiv.org/abs/1809.06923} {arXiv:1809.06923 [astro-ph.CO]} \BibitemShut {NoStop}%
\bibitem [{\citenamefont {Rajantie}\ and\ \citenamefont {Stopyra}(2017)}]{Rajantie:2016hkj}%
  \BibitemOpen
  \bibfield  {author} {\bibinfo {author} {\bibfnamefont {A.}~\bibnamefont {Rajantie}}\ and\ \bibinfo {author} {\bibfnamefont {S.}~\bibnamefont {Stopyra}},\ }\href {\doibase 10.1103/PhysRevD.95.025008} {\bibfield  {journal} {\bibinfo  {journal} {Phys. Rev. D}\ }\textbf {\bibinfo {volume} {95}},\ \bibinfo {pages} {025008} (\bibinfo {year} {2017})},\ \Eprint {http://arxiv.org/abs/1606.00849} {arXiv:1606.00849 [hep-th]} \BibitemShut {NoStop}%
\bibitem [{\citenamefont {Espinosa}(2020)}]{Espinosa:2020qtq}%
  \BibitemOpen
  \bibfield  {author} {\bibinfo {author} {\bibfnamefont {J.~R.}\ \bibnamefont {Espinosa}},\ }\href {\doibase 10.1088/1475-7516/2020/06/052} {\bibfield  {journal} {\bibinfo  {journal} {JCAP}\ }\textbf {\bibinfo {volume} {06}},\ \bibinfo {pages} {052} (\bibinfo {year} {2020})},\ \Eprint {http://arxiv.org/abs/2003.06219} {arXiv:2003.06219 [hep-ph]} \BibitemShut {NoStop}%
\bibitem [{\citenamefont {Gialamas}\ \emph {et~al.}(2023)\citenamefont {Gialamas}, \citenamefont {Karam},\ and\ \citenamefont {Pappas}}]{Gialamas:2022gxv}%
  \BibitemOpen
  \bibfield  {author} {\bibinfo {author} {\bibfnamefont {I.~D.}\ \bibnamefont {Gialamas}}, \bibinfo {author} {\bibfnamefont {A.}~\bibnamefont {Karam}}, \ and\ \bibinfo {author} {\bibfnamefont {T.~D.}\ \bibnamefont {Pappas}},\ }\href {\doibase 10.1016/j.physletb.2023.137885} {\bibfield  {journal} {\bibinfo  {journal} {Phys. Lett. B}\ }\textbf {\bibinfo {volume} {840}},\ \bibinfo {pages} {137885} (\bibinfo {year} {2023})},\ \Eprint {http://arxiv.org/abs/2212.03052} {arXiv:2212.03052 [hep-ph]} \BibitemShut {NoStop}%
\bibitem [{\citenamefont {Gialamas}\ and\ \citenamefont {Veerm\"ae}(2023)}]{Gialamas:2023emn}%
  \BibitemOpen
  \bibfield  {author} {\bibinfo {author} {\bibfnamefont {I.~D.}\ \bibnamefont {Gialamas}}\ and\ \bibinfo {author} {\bibfnamefont {H.}~\bibnamefont {Veerm\"ae}},\ }\href {\doibase 10.1016/j.physletb.2023.138109} {\bibfield  {journal} {\bibinfo  {journal} {Phys. Lett. B}\ }\textbf {\bibinfo {volume} {844}},\ \bibinfo {pages} {138109} (\bibinfo {year} {2023})},\ \Eprint {http://arxiv.org/abs/2305.07693} {arXiv:2305.07693 [hep-th]} \BibitemShut {NoStop}%
\bibitem [{\citenamefont {McDonald}(2024)}]{McDonald:2023qzz}%
  \BibitemOpen
  \bibfield  {author} {\bibinfo {author} {\bibfnamefont {J.}~\bibnamefont {McDonald}},\ }\href {\doibase 10.1088/1475-7516/2024/10/096} {\bibfield  {journal} {\bibinfo  {journal} {JCAP}\ }\textbf {\bibinfo {volume} {10}},\ \bibinfo {pages} {096} (\bibinfo {year} {2024})},\ \Eprint {http://arxiv.org/abs/2305.16208} {arXiv:2305.16208 [hep-ph]} \BibitemShut {NoStop}%
\bibitem [{\citenamefont {Garcia-Bellido}\ \emph {et~al.}(2009)\citenamefont {Garcia-Bellido}, \citenamefont {Figueroa},\ and\ \citenamefont {Rubio}}]{Garcia-Bellido:2008ycs}%
  \BibitemOpen
  \bibfield  {author} {\bibinfo {author} {\bibfnamefont {J.}~\bibnamefont {Garcia-Bellido}}, \bibinfo {author} {\bibfnamefont {D.~G.}\ \bibnamefont {Figueroa}}, \ and\ \bibinfo {author} {\bibfnamefont {J.}~\bibnamefont {Rubio}},\ }\href {\doibase 10.1103/PhysRevD.79.063531} {\bibfield  {journal} {\bibinfo  {journal} {Phys. Rev. D}\ }\textbf {\bibinfo {volume} {79}},\ \bibinfo {pages} {063531} (\bibinfo {year} {2009})},\ \Eprint {http://arxiv.org/abs/0812.4624} {arXiv:0812.4624 [hep-ph]} \BibitemShut {NoStop}%
\bibitem [{\citenamefont {Bezrukov}\ \emph {et~al.}(2009)\citenamefont {Bezrukov}, \citenamefont {Magnin},\ and\ \citenamefont {Shaposhnikov}}]{Bezrukov:2008ej}%
  \BibitemOpen
  \bibfield  {author} {\bibinfo {author} {\bibfnamefont {F.~L.}\ \bibnamefont {Bezrukov}}, \bibinfo {author} {\bibfnamefont {A.}~\bibnamefont {Magnin}}, \ and\ \bibinfo {author} {\bibfnamefont {M.}~\bibnamefont {Shaposhnikov}},\ }\href {\doibase 10.1016/j.physletb.2009.03.035} {\bibfield  {journal} {\bibinfo  {journal} {Phys. Lett. B}\ }\textbf {\bibinfo {volume} {675}},\ \bibinfo {pages} {88} (\bibinfo {year} {2009})},\ \Eprint {http://arxiv.org/abs/0812.4950} {arXiv:0812.4950 [hep-ph]} \BibitemShut {NoStop}%
\bibitem [{\citenamefont {Barvinsky}\ \emph {et~al.}(2008)\citenamefont {Barvinsky}, \citenamefont {Kamenshchik},\ and\ \citenamefont {Starobinsky}}]{Barvinsky:2008ia}%
  \BibitemOpen
  \bibfield  {author} {\bibinfo {author} {\bibfnamefont {A.~O.}\ \bibnamefont {Barvinsky}}, \bibinfo {author} {\bibfnamefont {A.~Y.}\ \bibnamefont {Kamenshchik}}, \ and\ \bibinfo {author} {\bibfnamefont {A.~A.}\ \bibnamefont {Starobinsky}},\ }\href {\doibase 10.1088/1475-7516/2008/11/021} {\bibfield  {journal} {\bibinfo  {journal} {JCAP}\ }\textbf {\bibinfo {volume} {11}},\ \bibinfo {pages} {021} (\bibinfo {year} {2008})},\ \Eprint {http://arxiv.org/abs/0809.2104} {arXiv:0809.2104 [hep-ph]} \BibitemShut {NoStop}%
\bibitem [{\citenamefont {Shaposhnikov}\ and\ \citenamefont {Zenhausern}(2009)}]{Shaposhnikov:2008xi}%
  \BibitemOpen
  \bibfield  {author} {\bibinfo {author} {\bibfnamefont {M.}~\bibnamefont {Shaposhnikov}}\ and\ \bibinfo {author} {\bibfnamefont {D.}~\bibnamefont {Zenhausern}},\ }\href {\doibase 10.1016/j.physletb.2008.11.041} {\bibfield  {journal} {\bibinfo  {journal} {Phys. Lett. B}\ }\textbf {\bibinfo {volume} {671}},\ \bibinfo {pages} {162} (\bibinfo {year} {2009})},\ \Eprint {http://arxiv.org/abs/0809.3406} {arXiv:0809.3406 [hep-th]} \BibitemShut {NoStop}%
\bibitem [{\citenamefont {De~Simone}\ \emph {et~al.}(2009)\citenamefont {De~Simone}, \citenamefont {Hertzberg},\ and\ \citenamefont {Wilczek}}]{DeSimone:2008ei}%
  \BibitemOpen
  \bibfield  {author} {\bibinfo {author} {\bibfnamefont {A.}~\bibnamefont {De~Simone}}, \bibinfo {author} {\bibfnamefont {M.~P.}\ \bibnamefont {Hertzberg}}, \ and\ \bibinfo {author} {\bibfnamefont {F.}~\bibnamefont {Wilczek}},\ }\href {\doibase 10.1016/j.physletb.2009.05.054} {\bibfield  {journal} {\bibinfo  {journal} {Phys. Lett. B}\ }\textbf {\bibinfo {volume} {678}},\ \bibinfo {pages} {1} (\bibinfo {year} {2009})},\ \Eprint {http://arxiv.org/abs/0812.4946} {arXiv:0812.4946 [hep-ph]} \BibitemShut {NoStop}%
\bibitem [{\citenamefont {Barvinsky}\ \emph {et~al.}(2009)\citenamefont {Barvinsky}, \citenamefont {Kamenshchik}, \citenamefont {Kiefer}, \citenamefont {Starobinsky},\ and\ \citenamefont {Steinwachs}}]{Barvinsky:2009fy}%
  \BibitemOpen
  \bibfield  {author} {\bibinfo {author} {\bibfnamefont {A.~O.}\ \bibnamefont {Barvinsky}}, \bibinfo {author} {\bibfnamefont {A.~Y.}\ \bibnamefont {Kamenshchik}}, \bibinfo {author} {\bibfnamefont {C.}~\bibnamefont {Kiefer}}, \bibinfo {author} {\bibfnamefont {A.~A.}\ \bibnamefont {Starobinsky}}, \ and\ \bibinfo {author} {\bibfnamefont {C.}~\bibnamefont {Steinwachs}},\ }\href {\doibase 10.1088/1475-7516/2009/12/003} {\bibfield  {journal} {\bibinfo  {journal} {JCAP}\ }\textbf {\bibinfo {volume} {12}},\ \bibinfo {pages} {003} (\bibinfo {year} {2009})},\ \Eprint {http://arxiv.org/abs/0904.1698} {arXiv:0904.1698 [hep-ph]} \BibitemShut {NoStop}%
\bibitem [{\citenamefont {Barvinsky}\ \emph {et~al.}(2012)\citenamefont {Barvinsky}, \citenamefont {Kamenshchik}, \citenamefont {Kiefer}, \citenamefont {Starobinsky},\ and\ \citenamefont {Steinwachs}}]{Barvinsky:2009ii}%
  \BibitemOpen
  \bibfield  {author} {\bibinfo {author} {\bibfnamefont {A.~O.}\ \bibnamefont {Barvinsky}}, \bibinfo {author} {\bibfnamefont {A.~Y.}\ \bibnamefont {Kamenshchik}}, \bibinfo {author} {\bibfnamefont {C.}~\bibnamefont {Kiefer}}, \bibinfo {author} {\bibfnamefont {A.~A.}\ \bibnamefont {Starobinsky}}, \ and\ \bibinfo {author} {\bibfnamefont {C.~F.}\ \bibnamefont {Steinwachs}},\ }\href {\doibase 10.1140/epjc/s10052-012-2219-3} {\bibfield  {journal} {\bibinfo  {journal} {Eur. Phys. J. C}\ }\textbf {\bibinfo {volume} {72}},\ \bibinfo {pages} {2219} (\bibinfo {year} {2012})},\ \Eprint {http://arxiv.org/abs/0910.1041} {arXiv:0910.1041 [hep-ph]} \BibitemShut {NoStop}%
\bibitem [{\citenamefont {Markkanen}\ \emph {et~al.}(2018{\natexlab{b}})\citenamefont {Markkanen}, \citenamefont {Nurmi}, \citenamefont {Rajantie},\ and\ \citenamefont {Stopyra}}]{Markkanen:2018bfx}%
  \BibitemOpen
  \bibfield  {author} {\bibinfo {author} {\bibfnamefont {T.}~\bibnamefont {Markkanen}}, \bibinfo {author} {\bibfnamefont {S.}~\bibnamefont {Nurmi}}, \bibinfo {author} {\bibfnamefont {A.}~\bibnamefont {Rajantie}}, \ and\ \bibinfo {author} {\bibfnamefont {S.}~\bibnamefont {Stopyra}},\ }\href {\doibase 10.1007/JHEP06(2018)040} {\bibfield  {journal} {\bibinfo  {journal} {JHEP}\ }\textbf {\bibinfo {volume} {06}},\ \bibinfo {pages} {040} (\bibinfo {year} {2018}{\natexlab{b}})},\ \Eprint {http://arxiv.org/abs/1804.02020} {arXiv:1804.02020 [hep-ph]} \BibitemShut {NoStop}%
\bibitem [{\citenamefont {Nielsen}(1975)}]{Nielsen:1975fs}%
  \BibitemOpen
  \bibfield  {author} {\bibinfo {author} {\bibfnamefont {N.~K.}\ \bibnamefont {Nielsen}},\ }\href {\doibase 10.1016/0550-3213(75)90301-6} {\bibfield  {journal} {\bibinfo  {journal} {Nucl. Phys. B}\ }\textbf {\bibinfo {volume} {101}},\ \bibinfo {pages} {173} (\bibinfo {year} {1975})}\BibitemShut {NoStop}%
\bibitem [{\citenamefont {'t~Hooft}\ and\ \citenamefont {Veltman}(1974)}]{tHooft:1974toh}%
  \BibitemOpen
  \bibfield  {author} {\bibinfo {author} {\bibfnamefont {G.}~\bibnamefont {'t~Hooft}}\ and\ \bibinfo {author} {\bibfnamefont {M.~J.~G.}\ \bibnamefont {Veltman}},\ }\href@noop {} {\bibfield  {journal} {\bibinfo  {journal} {Ann. Inst. H. Poincare A Phys. Theor.}\ }\textbf {\bibinfo {volume} {20}},\ \bibinfo {pages} {69} (\bibinfo {year} {1974})}\BibitemShut {NoStop}%
\bibitem [{\citenamefont {Deser}\ and\ \citenamefont {van Nieuwenhuizen}(1974{\natexlab{a}})}]{Deser:1974cy}%
  \BibitemOpen
  \bibfield  {author} {\bibinfo {author} {\bibfnamefont {S.}~\bibnamefont {Deser}}\ and\ \bibinfo {author} {\bibfnamefont {P.}~\bibnamefont {van Nieuwenhuizen}},\ }\href {\doibase 10.1103/PhysRevD.10.411} {\bibfield  {journal} {\bibinfo  {journal} {Phys. Rev. D}\ }\textbf {\bibinfo {volume} {10}},\ \bibinfo {pages} {411} (\bibinfo {year} {1974}{\natexlab{a}})}\BibitemShut {NoStop}%
\bibitem [{\citenamefont {Deser}\ and\ \citenamefont {van Nieuwenhuizen}(1974{\natexlab{b}})}]{Deser:1974cz}%
  \BibitemOpen
  \bibfield  {author} {\bibinfo {author} {\bibfnamefont {S.}~\bibnamefont {Deser}}\ and\ \bibinfo {author} {\bibfnamefont {P.}~\bibnamefont {van Nieuwenhuizen}},\ }\href {\doibase 10.1103/PhysRevD.10.401} {\bibfield  {journal} {\bibinfo  {journal} {Phys. Rev. D}\ }\textbf {\bibinfo {volume} {10}},\ \bibinfo {pages} {401} (\bibinfo {year} {1974}{\natexlab{b}})}\BibitemShut {NoStop}%
\bibitem [{\citenamefont {Deser}\ \emph {et~al.}(1974)\citenamefont {Deser}, \citenamefont {Tsao},\ and\ \citenamefont {van Nieuwenhuizen}}]{Deser:1974xq}%
  \BibitemOpen
  \bibfield  {author} {\bibinfo {author} {\bibfnamefont {S.}~\bibnamefont {Deser}}, \bibinfo {author} {\bibfnamefont {H.-S.}\ \bibnamefont {Tsao}}, \ and\ \bibinfo {author} {\bibfnamefont {P.}~\bibnamefont {van Nieuwenhuizen}},\ }\href {\doibase 10.1103/PhysRevD.10.3337} {\bibfield  {journal} {\bibinfo  {journal} {Phys. Rev. D}\ }\textbf {\bibinfo {volume} {10}},\ \bibinfo {pages} {3337} (\bibinfo {year} {1974})}\BibitemShut {NoStop}%
\bibitem [{\citenamefont {Parker}\ and\ \citenamefont {Toms}(2009)}]{Parker:2009uva}%
  \BibitemOpen
  \bibfield  {author} {\bibinfo {author} {\bibfnamefont {L.~E.}\ \bibnamefont {Parker}}\ and\ \bibinfo {author} {\bibfnamefont {D.}~\bibnamefont {Toms}},\ }\href {\doibase 10.1017/CBO9780511813924} {\emph {\bibinfo {title} {{Quantum Field Theory in Curved Spacetime}: {Quantized Field and Gravity}}}},\ Cambridge Monographs on Mathematical Physics\ (\bibinfo  {publisher} {Cambridge University Press},\ \bibinfo {year} {2009})\BibitemShut {NoStop}%
\bibitem [{\citenamefont {Bezrukov}\ and\ \citenamefont {Shaposhnikov}(2009)}]{Bezrukov:2009db}%
  \BibitemOpen
  \bibfield  {author} {\bibinfo {author} {\bibfnamefont {F.}~\bibnamefont {Bezrukov}}\ and\ \bibinfo {author} {\bibfnamefont {M.}~\bibnamefont {Shaposhnikov}},\ }\href {\doibase 10.1088/1126-6708/2009/07/089} {\bibfield  {journal} {\bibinfo  {journal} {JHEP}\ }\textbf {\bibinfo {volume} {07}},\ \bibinfo {pages} {089} (\bibinfo {year} {2009})},\ \Eprint {http://arxiv.org/abs/0904.1537} {arXiv:0904.1537 [hep-ph]} \BibitemShut {NoStop}%
\end{thebibliography}%
\clearpage
\newpage
\maketitle
\onecolumngrid
\begin{center}
\textbf{\large Magnetic Anti-de Sitter wormholes as seeds for Higgs inflation} 
\\ 
\vspace{0.05in}
{Panos Betzios, Ioannis D. Gialamas and Olga Papadoulaki}
\\ 
\vspace{0.07in}
{ \it \large \hypertarget{sup_mat}{Supplemental Material}}
\end{center}
\onecolumngrid

In this supplemental material, we provide some further details on the scales and properties of Higgs inflation that are relevant for our work.

\section*{Scales}

Here we list some scales that are relevant in Higgs inflation.

\begin{itemize}

\item The Planck mass $\mpl \equiv 1/\sqrt{8 \pi G_N} = 2.435 \times 10^{18} \,{\rm GeV}$ and the electroweak scale Higgs vacuum expectation value (VEV) $v_{\rm ew} \simeq 250 \,{\rm GeV}$ (responsible for the generation of masses of the SM  particles).

    \item According to~\cite{ParticleDataGroup:2024cfk} the average value of the Higgs mass is $m_{H} = 125.20 \pm 0.11 \,{\rm GeV}$, while the top quark mass computed from direct measurements is $ m_t =172.57 \pm 0.29 \,{\rm GeV}$. The stability is very sensitive to $m_t$. The Higgs self coupling $\lambda(\mu)$ following from SM RG equations decreases up to some scale and then increases. Whether it remains positive or not depends
    on the Higgs mass and the top Yukawa coupling $y_t$ at each RG scale $\mu$. Within the present uncertainties there is a case of both stability, metastability or instability~\cite{Masina:2012tz,Espinosa:2015qea,Bednyakov:2015sca}. Based on the central values provided by~\cite{ParticleDataGroup:2024cfk} we are in the metastable region. Gravitational effects on vacuum stability were initially studied in~\cite{Coleman:1980aw}, while the impact of gravity on the stability of the SM  Higgs has been discussed in~\cite{Isidori:2007vm, Salvio:2016mvj, Branchina:2016bws, Markkanen:2018pdo, Rajantie:2016hkj, Espinosa:2020qtq}, as well as in alternative formulations of gravity in~\cite{Gialamas:2022gxv, Gialamas:2023emn}.

    \item We also have the following cutoff scales for the effective field theory of Higgs inflation: The higher cutoff is the Planck scale $\Lambda_{\rm UV} \leq \mpl$. The value of the cutoff in general depends on the VEV/value of the Higgs boson $\phi$. Above the EW vacuum the cutoff is~\cite{Burgess:2009ea,Barbon:2009ya} $\Lambda_{ \rm ew} = \mpl /\xi$. In the inflationary regime the cutoff is $\Lambda_{\rm inf} = \mpl / \sqrt{\xi}$, which for values of $\xi < 10^3$ that we are interested in, is much higher than the energies of the processes of interest. In the case of metastable Higgs inflation the value of $\xi$ can be even lower than that~\cite{Bezrukov:2014ipa} (see also~\cite{McDonald:2023qzz}), raising the value of the cutoff even further, giving further credence to our scenario. 
    
\end{itemize}

\section*{Einstein frame for Higgs inflation and inflationary observables}

The scalar field part of the action~(1) of the main text can be expressed in the Einstein frame through a Weyl rescaling of the form
\be
\label{eq:Weyl_ag}
g_{\m\n} \rightarrow \Omega^{-2}(\phi) g_{\m\n}\,, \qquad \text{with} \qquad \Omega^{2}(\phi) = 1 +\k\xi\phi^2\,.
\ee
So,
\be
\mathcal{S}_E = \int {\rm d}^4x\sqrt{g_E} \left(-\frac{R}{2\k} +\left( \frac{1+(1+6\xi)\k\xi\phi^2}{(1+\k\xi\phi^2)^2}\right)\frac{\partial_\m\phi\partial^\m\phi}{2} +\frac{V(\phi)}{(1+\k\xi\phi^2)^2}  \right)\,.
\ee
Given that the kinetic term is not in a canonical form, we perform the following field redefinition,
\be
{\rm d}\chi = \sqrt{\frac{1+(1+6\xi)\k\xi\phi^2}{(1+\k\xi\phi^2)^2}} {\rm d}\phi\,.
\ee
Integrating the previous equation we obtain~\cite{Garcia-Bellido:2008ycs}
\be
\chi(\phi) = \frac{1}{\sqrt{\k\xi}} \left[\sqrt{1+6\xi}\sinh^{-1} (\sqrt{1+6\xi}u) -\sqrt{6\xi}\sinh^{-1} \left(\sqrt{6\xi}\frac{u}{\sqrt{1+u^2}} \right) \right]\,,
\ee
where $u=\sqrt{\k\xi}\phi$. The above can be approximated as
\be
\chi\simeq \begin{cases}
   \displaystyle    \phi\,, & \text{if }\,\, \phi\ll \mpl/\xi\\[0.3cm]
   \displaystyle     \sqrt{\frac{3}{2\k}} \log[1+\k\xi\phi^2]\,, & \text{if }\, \, \phi\gg  \mpl/\xi\,.
    \end{cases}
\ee
So the Einstein frame action is
\be
\mathcal{S}_E = \int {\rm d}^4x\sqrt{g_E} \left(-\frac{R}{2\k} +\frac{\partial_\m\chi\partial^\m\chi}{2} +U(\chi)  \right)\,,
\ee
with
\be
U(\chi) = \frac{V(\phi(\chi))}{(1+\k\xi\phi^2(\chi))^2} \simeq \begin{cases}
   \displaystyle    \frac{\l}{4}(\chi^2-v_{\rm ew}^2)^2\,, & \text{if }\,\, \chi\ll \mpl/\xi\\[0.3cm]
   \displaystyle \frac{\l}{4\k^2\xi^2}\left(1-e^{-\sqrt{2\k/3}\chi} \right)^2 \,, & \text{if }\, \, \chi\gg  \mpl/\xi\,,
    \end{cases}
\ee
where we used the standard (tree-level) Higgs potential
$V(\phi) = \lambda (\phi^2 - v_{\rm ew}^2)^2/4 $ and assumed that $\xi \ll \mpl^2/v_{\rm ew}^2$.

In this model at tree level, the inflationary observables (for $N_\star \gg 1$)  come out to be~\cite{Bezrukov:2007ep}
\be
A^\star_s = \frac{N_\star^2 \lambda}{72 \pi^2 \xi^2} \, , \qquad n_s^\star = 1 - \frac{2}{N_\star} \approx 0.9667 \, , \qquad r_\star = \frac{12}{N_\star^2} \approx 0.0033 \, ,
\ee
where the approximate values are for $N_\star = 60$ $e$-folds of inflation. In the main text the tree level couplings are replaced by running couplings that depend on the RG-scale $\mu$, and evaluated at the scale of Inflation $\mu_\star$.

Observations of the cosmic microwave background (CMB) place significant constraints on inflationary predictions, as shown in~\cite{Planck:2018jri,BICEP:2021xfz}. The latest combined data from Planck, BICEP/Keck, and BAO have set the following limits on observable values at the pivot scale $k_\star = 0.05\, {\rm Mpc}^{-1}$:
\begin{align}
\label{eq:planck_bounds}
{A}^\star_s = (2.10 \pm 0.03) \times 10^{-9} & \, , \qquad 68\%\,  \text{CL} \nonumber  \\
   {n}_s^\star = 0.9649 \pm 0.0042& \, , \qquad 68\%\, \text{CL}  \\
   r_\star < 0.036& \, , \qquad 95\%\, \text{CL} \, . \nonumber
\end{align}

These constrain the ratio $\lambda_\star/\xi^2_\star \sim 5 \times 10^{-10}$ to be small. Notice that in our setup, where the couplings do run this constraint is not a constraint in the values of the couplings in our EW scale $\mu_{\rm ew}$, but a constraint on their value at the inflationary scale $\mu_\star$. In particular if $\lambda_\star$ is suffiently small at that scale (for higher energies it turns negative), $\xi_\star$ can also be  much smaller than in the tree level scenario. Their exact values though are very sensitive to the precise values of the Higgs and Top quark masses at the EW scale and the RG running to the inflationary scale.

 \section*{Renormalization group in various schemes}

An analysis of Higgs non-minimal $\xi$-inflation must also include the renormalization group (RG) running of the couplings and loop corrections to the (effective) action. Under RG, the minimal set of higher-dimensional operators to be included on top of the tree-level action is generated by the theory itself via radiative corrections. Since our model of Einstein gravity coupled to the SM  constitutes a non-renormalizable theory, the cancellation of the loop divergences stemming from the original action requires the inclusion of an infinite set of counterterms. We can therefore threat this model only as an effective field theory (EFT). The outcome of the subtraction procedure of divergences depends on the renormalization scheme, with different choices corresponding to different assumptions about the ultraviolet (UV) completion of the model. In any RG-scheme, the best prescription is to chose a scale $\mu$ such that the effect of the logarithms is minimised as possible (so in general the scale will also depend on the background field values i.e. $\mu(\phi)$). Moreover this choice of scheme is not frame independent, but transforms under Weyl transformations \eqref{eq:Weyl_ag} as $\mu \rightarrow \mu/\Omega$. 
In our context of Higgs inflation, there exist two basic inequivalent renormalization prescriptions/schemes depending on whether quantum corrections to the potential and effective action are computed setting a fixed, field independent cutoff scale in the Einstein frame (called prescription I)~\cite{Bezrukov:2007ep,Bezrukov:2008ej,Bezrukov:2014ipa}, or in the Jordan frame (called prescription II)~\cite{Barvinsky:2008ia,Shaposhnikov:2008xi,Barvinsky:2008ia,DeSimone:2008ei,Barvinsky:2009fy,Barvinsky:2009ii}. These correspond to different UV-completions of the model. The first prescription is connected to the idea of obtaining quantum scale invariance at high energies, while the second is motivated by using the standard RG prescription for computing loop corrections and measuring physical distances and scales with respect to a fixed Planck cutoff. 

We should also add at this point that there exists a third prescription (III)~\cite{Markkanen:2018bfx} in which the RG-scale $\mu$ depends also on the curvature $R$, which can potentially become important if the size of the early universe is sufficiently small and approaches the Planck scale. In the next section, we prove that in our model it is safe to neglect any such curvature dependence to define the renormalization point/substraction scheme, since the relevant wormhole backgrounds that we consider have a curvature scale that is found to be at least nine orders smaller than $\mpl^2$.
 
The discrepancy between (I)-(II) becomes even more pronounced at the scale where $\lambda(\mu)$ becomes small or negative. This is important since as we argued, central values favor Higgs metastability, and there exists a window of scales where $\lambda(\mu) < 0$. The inflationary scale $\mu_\star$ can be lower or higher than this window, and this is highly dependent on the value of the mass of the top quark and the resulting RG running, which we inevitably need to take into account. We are thus led to understand the properties of the loop corrected RG improved effective potential. Our preferred scheme choice will be the one that minimizes the running logarithms, and in particular at the inflationary scale $\mu_\star$.
This is the preferred scheme choice from an EFT perspective, since it does not include any assumptions about the UV completion of the EFT model.

Our perspective hence, is that both the non-minimal coupling as well as higher derivative terms in the effective action are induced by quantum loop corrections. We are able to show that higher derivative terms are highly supressed for the backgrounds and regimes of physical interest. Therefore we minimize the logarithms and renormalize our model in the original frame where the loop corrections are computed. Finally we transform the resulting action and effective potential into the Einstein frame (the Weyl transformation \eqref{eq:Weyl_ag} becomes scale dependent, since the couplings themselves depend on the scale $\mu$).

\section*{RG Improved effective potential}

We define the Callan-Symanzik equation governing the running of the effective potential
\be\label{CSequation}
\left(\frac{\partial}{\partial \mu} + \sum_i \beta_i(g_i) \frac{\partial}{\partial g_i} + \gamma_\phi \phi \frac{\partial}{\partial \phi} \right) V_{\rm RGE}(\phi ; g_i ; \mu) = 0 \, ,
\ee
in terms of the running couplings $g_i(\mu)$ (and the corresponding beta-functions)
\be\label{Betafunctions}
\frac{{\rm d}  g_i}{{\rm d} \log \mu} = \beta_i(g_i ; \mu) \, , \qquad \frac{{\rm d}  \phi(\mu)}{{\rm d} \log \mu} = \gamma_\phi \phi(\mu) \, ,
\ee
and initial conditions at $\mu_0$. The renormalization point is chosen such that one minimizes the logarithms as much as possible. At this point, and in line to our previous discussion, one should keep in mind that the precise shape of the effective potential is gauge and RG scheme dependent. Nevertheless both the extrema and the value of the potential at them are gauge independent as demanded by the Nielsen identity~\cite{Nielsen:1975fs} (this provides us with crucial information about various physical couplings and observables at the inflationary plateau and electroweak scale). 

In the rest we choose a tree level Higgs potential
\be
V_{\text{tree}}(\phi) =  \frac{m^2}{2} \phi^2 + \frac{\lambda}{4} \phi^4 \, .
\ee 
For the SM, one defines
the quantities
\bea
Y_2 = 3(y_u^2 +y_c^2 + y_t^2) + 3 (y_d^2 + y_s^2 + y_b^2) + (y_e^2+y_\mu^2+y_\t^2) \, , \nn \\
Y_4 = 3(y_u^4 +y_c^4 + y_t^4) + 3 (y_d^4 + y_s^4 + y_b^4) + (y_e^4+y_\mu^4+y_\t^4) \, ,
\eea
in terms of Yukawa couplings ($y_i$) for each fermion of the SM. The SM beta functions at one-loop are then given by~\cite{Buttazzo:2013uya}
\begin{align}\label{eq:betas}
{16\pi^2}\beta_{y_t} &= {y_t}\left[\frac{3}{2}(y_t^2 - y_b^2) + Y_2 - \left(\frac{17}{12}(g')^2 + \frac{9}{4}g^2 + 8g_3^2\right) \right] \, , \nn \\
{16\pi^2}\beta_{y_b} &= {y_b}\left[\frac{3}{2}(y_b^2 - y_t^2) + Y_2 - \left(\frac{5}{12}(g')^2 + \frac{9}{4}g^2 + 8g_3^2\right) \right] \, , \nn \\
{16\pi^2}\beta_{y_l} &= {y_l}\left[\frac{3}{2}y_l^2 + Y_2 - \left(\frac{45}{12}(g')^2 + \frac{9}{4}g^2\right) \right]\, , \nn \\
{16\pi^2}\beta_{\lambda} &=24\lambda^2 - 3\lambda \left((g')^2 + 3g^2\right) + \frac{3}{4}\left(\frac{1}{2}(g')^4 + (g')^2g^2 + \frac{3}{2}g^4\right) + 4Y_2\lambda - 2Y_4\, , \nn \\
{16\pi^2}\beta_{m^2} &={m^2}\left[12\lambda - \frac{3}{2}(g')^2 - \frac{9}{2}g^2 + 2Y_2\right]\,, \nn \\
{16\pi^2}\beta_{g'} &= \frac{41}{6}{(g')^3}\, , \qquad
{16\pi^2}\beta_{g} =  -\frac{19}{6}{g^3},\qquad
{16\pi^2}\beta_{g_3} = -7 {g_3^4}\, ,
\end{align}
where $\beta_{y_l}$ is each lepton beta function, $g'$ is the U$(1)$ coupling, 
$g$ is the SU$(2)$ coupling, and $g_3$ the SU$(3)$ coupling. At one-loop order, we can neglect the running of $\zeta_W$ and $\zeta_Z$ (it enters at two loops). The field renormalization of $\phi$ (anomalous dimension) is found to be
\be
\gamma_\phi = \frac{1}{16\pi^2}\bigg[
Y_2-\frac{9 g^2}{4}-\frac{3(g')^2}{4}-\zeta_W\frac{ g^2}{2}-\zeta_Z\frac{1}{4}\left(g^2+(g')^2\right)\bigg] \, .
\ee
When including gravity at one-loop order, one needs to include the following additional terms in the purely gravitational part of the Euclidean effective action~\cite{tHooft:1974toh}
\be
S_{GR}^E = \int {\rm d}^4 x \sqrt{g_E} \left( - \kappa_1 R + \alpha_1 R^2 + \alpha_2 R_{\m \n} R^{\m \n} + \alpha_3 R_{\m \n \rho \sigma} R^{\m \n \rho \sigma} \right) \, ,
\ee
together with the non-minimal coupling of our interest
\be
S_\xi = - \int {\rm d}^4 x \sqrt{g_E} \frac{\xi}{2} \phi^2 R \, .
\ee
Collecting the contributions of the various SM fields, the gravitational part of the $\beta$-functions is (this set together with eqns.~\eqref{eq:betas} is exhaustive to one-loop order~\cite{Markkanen:2018bfx}, see also~\cite{Deser:1974cy,Deser:1974cz,Deser:1974xq}, for results using dimensional regularization)
\begin{subequations}
\begin{align}\label{eq:GRbeta}
16\pi^2 \beta_{\xi} &= \bigg(\xi - \frac{1}{6}\bigg)\bigg[12 \lambda +
2Y_2-\frac{3 (g')^2}{2}-\frac{9 g^2}{2}\bigg]\,, \\ {16\pi^2}\beta_{\kappa_1}&={4}{m^2\bigg(\xi - \frac{1}{6}\bigg)}\,, \\ {16\pi^2}\beta_{\alpha_1}&=2 \xi ^2-\frac{2 \xi }{3}-\frac{277}{144}\,,  \\ 
16\pi^2 \beta_{\alpha_2}&=\frac{571}{90} \,, \\
16\pi^2 \beta_{\alpha_3}&=-\frac{293}{720} \,.
\end{align}
\end{subequations}
Notice though that in four dimensions  there is an identity
\be
\sqrt{g} \left(R_{\m \n \rho \sigma} R^{\m \n \rho \sigma} - 4 R^{\m \n} R_{\m \n} + R^2 \right) = \text{total derivative} \, , 
\ee
and hence one can equivalently neglect the 
$\alpha_3$ coupling by replacing
\be
\alpha_2 \rightarrow \alpha_2 + 4 {\alpha_3}  \, , \quad \alpha_1 \rightarrow \alpha_1 - \alpha_3\,.
\ee
The resulting beta functions reveal that generically the gravitational higher derivative couplings decrease in the IR, while they increase and hit some Landau pole in the UV, that defines a scale above which the effective action ceases to be valid. For our purposes we shall find that it is consistent to neglect the higher derivative terms in the low-energy effective action, keeping only the non-minimal coupling term. Due to the form of its beta function in \eqref{eq:GRbeta},
this non-minimal coupling $\xi$, will generically run unless $\xi = 1/6$ (conformal coupling point).

Assuming that quantum fluctuations of the fields do not change the symmetry of the background metric, in our FLRW type of background~(eq.~(7) in the main text), the quadratic in curvature terms combine to give
\begin{align}
\mathcal{S} &= \text{Vol}_{S^3} \int {\rm d} \tau \frac{12}{a}  \bigg[\tilde{\a}_1\left(a^2(a'')^2+(a'^2-1)^2\right) +\tilde{\a}_2 (a'^2-1)a a'' \bigg] \nn \\
&= \text{Vol}_{S^3} \int {\rm d} \tau \frac{12}{a}  \bigg[\tilde{\a}_1\left(a^2(a'')^2+(a'^2-1)^2\right) - \tilde{\a}_2 a a'' \bigg] \, + \, \text{total derivative} \, ,
\end{align}
where 
\be
\tilde{\a}_1 = 3\a_1+\a_2+\a_3 \qquad \text{and} \qquad \tilde{\a}_2 = 6\a_1+\a_2 - 2 a_3 \,.
\ee
The corresponding beta functions read
\be
16\pi^2\b_{\tilde{\a}_1} =6 \left(\xi-\frac{1}{6}\right)^2 \geq 0 \qquad \text{and} \qquad 16\pi^2\b_{\tilde{\a}_2} = 12\xi^2 - 4\xi-\frac{263}{60}  \, .
\ee
The first is always positive (except at the conformal point) and shows the irrelevance of the higher derivative terms, that is irrespective of the precise SM content due to cancellations in the definition of $\tilde{a}_1$ and the FRLW ansatze. The second is positive for $\xi \gtrapprox 0.8 $ and negative for smaller $\xi$ (and can only affect the renormalization of the lower derivative terms). This proves that within our FLRW ansatze, the couplings affecting higher derivative terms will generically run to zero in the IR, leaving only the contribution of the non-minimal coupling term $\sim \xi \phi R$ together with the usual Einstein-Hilbert term and the SM Lagrangian.

We shall now describe the properties of the one-loop RGE improved effective potential $V_{\rm RGE}(\phi ; g_i ; \mu)$ for the scalar (Higgs) field. In order to do so it is convenient to introduce the Heat kernel coefficients~\cite{Birrell:1982ix,Parker:2009uva,Markkanen:2018bfx}
\bea\label{eq:heatkernel}
a_{2,s} &=& - \frac{1}{180} R_{\m \n} R^{\m \n} + \frac{1}{180} R_{\m \n \rho \sigma} R^{\m \n \rho \sigma} \, , \nn \\
{\rm Tr}(a_{2,f}) &=& {\rm Tr}_G (I) \left( - \frac{1}{45} R_{\m \n} R^{\m \n} - \frac{7}{360} R_{\m \n \rho \sigma} R^{\m \n \rho \sigma} \right) \, , \nn \\ 
\tns{(a_{2,g})}{^\a_\b} &=& - \frac{\tns{g}{^\a_\b}}{180} R_{\m \n} R^{\m \n} + \frac{\tns{g}{^\a_\b}}{180} R_{\m \n \rho \sigma} R^{\m \n \rho \sigma} + \frac{1}{12}\tns{R}{^\d_\m_\n_\b} \tns{R}{_\d^\m^\n^\a} \, ,
\eea
for the scalar, fermion and gauge fields respectively. The trace is over the corresponding group indices
(i.e. a $3$ for colour indices etc.). The terms in the effective potential that arise from one-loop contributions of the various SM fields exhibit the following two generic logarithmic behaviours
\be\label{term1}
n_i \mathcal{M}_i^4 \log \frac{|\mathcal{M}_i|^2}{\mu^2} \, ,
\ee
and
\be\label{term2}
a_{2, i} \log \frac{|\mathcal{M}_i|^2}{\mu^2} \, ,
\ee
with $a_{2, i}$ the various Heat-Kernel coefficients introduced above. In these formulae
the masses $\mathcal{M}_i$ are defined by
\be
\mathcal{M}_i^2(\mu ; \phi_{cl}) = \kappa_i(\mu) Z(\mu) \phi_{cl}^2 - \kappa_i'(\mu)  + \theta_i(\mu) R \, ,
\ee
in terms of coefficients that depend on the
running couplings of the theory (their explicit form can be found in~\cite{Markkanen:2018bfx}).

This result, leads one to chose a renormalization point
\be
\mu^2 = A \phi_{cl}^2 + B R + C \mpl^2 \, ,
\ee
with $A,B, C$ numerical coefficients, so that the logarithms will be as small as possible. This leads to an additional complication with aforementioned flat space schemes. Of course as the universe expands $R \rightarrow 0$ and one recovers the usual flat space renormalization schemes.

Nevertheless, in our model, when the scalar field is in the inflationary regime and above, its expectation value becomes very large: $\phi_{cl} \geq 2 \mpl/\sqrt{\xi_\star(1+6 \xi_\star)} \, , \, \xi_\star \ll 10^3$. If we seek wormhole backgrounds for which their characteristic size is always much larger than the Planck scale (macroscopic wormhole saddles), then to first order we can neglect the dependence of the curvature to the definition of the renormalization point. In fact, in our model we find that the throat curvature scale is set by $\tilde{U}_0 = \kappa U(0)/3$
\be
R_{throat}  \, \simeq \, 10 \, \tilde{U}_0 \,  \simeq  \, 4 \times 10^{-10} \mpl^2  \, ,
\ee
where in the last estimate, we used the condition that our model sustains $55-60$ $e$-folds of inflation, see the main text.
This indicates that the characteristic spatial size of our wormhole backgrounds that are relevant for phenomenology is always much greater than the Planck scale, giving further credence to our semi-classical approximations and allowing us to simplify the renormalization prescription using a ``small curvature expansion'' for the definition of the renormalization point and the one-loop effective action and scalar potential. 

In addition, one also finds that the terms of the type~\eqref{term1} dominate over the terms of the type~\eqref{term2}, by approximately eight orders of magnitude, and the effective potential is expressed to a very good approximation by
\bea
V^{\rm RG}_{\rm total}(\phi_{cl}) &=& V_{cl}(\phi_{cl}) + V^{\rm 1-loop}_{\rm total}(\phi_{cl}) \, , \nn \\
 V^{\rm 1-loop}_{\rm total}(\phi_{cl}) &\simeq&  \frac{1}{64 \pi^2} \sum_{i=1}^{31} \left( n_i \mathcal{M}_i^4(\mu) \left[ \log \left(\frac{|\mathcal{M}_i|^2(\mu)}{\mu^2} \right) - d_i \right] \right) \, ,
\eea
where $n_i, d_i$ depend on the SM field content, see the table \ref{tab:contributions} for fields that couple to the Higgs. There also exist fields that do not couple to Higgs to leading order in our small curvature expansion, and hence do not affect the effective potential and we can neglect their contribution.

\begin{table}
\centering  
\begin{tblr}{colspec  = {c|ccccc},
             colsep=6pt,
             row{1}   = {bg=gray!15} } 
				\hline\hline 
				$\Psi$ & $~~i$ & $~~n_i$   & $~d_i$    &$~~n'_i$    &$\mathcal{M}_i^2$    \\ \hline
				$~$ & $~~1$  & $~~2$       & $\quad{3}/{2}~~$        & $-34/15$        &  $m^2_W+R/12$      \\
				$~W^\pm$ & $~~2$  & $~~6$       & $\quad{5}/{6}~~$       & $-34/5$        &  $m^2_W+R/12$       \\
				$~$ & $~~3$  & $-2$      & $\quad{3}/{2}~~$         & $~~4/15$        & $ ~m^2_W-R/6$        \\\hline
				$~$ & $~~4$  & $~~1$        & $\quad{3}/{2}~~$ & $-17/15$        & $ m^2_Z+R/12$      \\
				$Z^0$ & $~~5$  & $~~3$        & $\quad{5}/{6}~~$ & $-17/5$        & $ m^2_Z+R/12$     \\
				$~$ & $~~6$  & $-1$      & $\quad{3}/{2}~~$ & $~~2/15$        & $ ~m^2_Z-R/6$     \\\hline
				q & $7-12$  & $-12$     & $\quad{3}/{2}~~$    & $~~38/5$        & $ m^2_{q}+R/12$      \\\hline
				$l$ & $13-15$  & $-4$     & $\quad{3}/{2}~~$    & $~38/15$        & $ m^2_{l}+R/12$      \\\hline
				$h$ & $~16$  & $~~1$       & $\quad{3}/{2}~~$          & $-2/15$      & $m_h^2+(\xi-{1}/{6})R$  \\\hline
				${\chi}_W$ & $~17$  & $~~2$      & $\quad{3}/{2}~~$           & $-4/15$      & $\,\,\,~~\quad m_\chi^2\,+\zeta_W m^2_W+(\xi-{1}/{6})R$   \\\hline
				${\chi}_Z$ & $~18$  & $~~1$       & $\quad{3}/{2}~~$           & $-2/15$      & $\,~~\quad m_\chi^2+\zeta_Z m^2_Z\,+(\xi-{1}/{6})R$   
				\\\hline ${c}_W$ & $~19$  & $-2$       & $\quad{3}/{2}~~$           & $~~4/15$      & $\zeta_W m^2_W-R/6$  
				\\\hline
				${c}_Z$ & $~20$  & $-1$      & $\quad{3}/{2}~~$            & $~~2/15$      & $\zeta_Z m^2_Z-R/6$   
				\\\hline
			\end{tblr}
   	\caption{\label{tab:contributions} The 1-loop contributions with tree-level couplings to the Higgs. $\Psi$ stands for $W^{\pm}$ and $Z^0$ bosons, the 6 quarks q, the 3 charged leptons $l$, the Higgs $h$. The Goldstone bosons are $\chi_{W}$ and $\chi_{Z}$ and ghosts $c_{W}$ and $c_{Z}$. The masses may be found~\cite{Markkanen:2018bfx}. (This table can also be found in~\cite{Markkanen:2018bfx}).}
\end{table}

Our analysis proves that in our model, up to a tremendous degree of accuracy ($\sim 10^{-8}$), that holds up to the inflationary scale and above, we can compute the loop corrected effective potential for the scalar Higgs using flat space RG results. This has already been performed in various works~\cite{Bezrukov:2009db,Barvinsky:2009fy}, using the type I and II schemes~\cite{Bezrukov:2007ep,Bezrukov:2008ej,Bezrukov:2014ipa,Barvinsky:2008ia,Shaposhnikov:2008xi,Barvinsky:2008ia,DeSimone:2008ei,Barvinsky:2009ii}. Our work shows that the best renormalization scheme from an EFT perspective is the one that reduces the logarithms as much as possible, so that in general $\mu^2 = A \phi_{cl}^2 + C \mpl^2$. In contrast with both schemes I, II in the literature, the precise numerical value of $A, C$ is properly fixed by minimizing logarithms at the scales of interest (i.e. inflationary scale $\mu_\star$), without the need to assume certain features for the UV completion of our EFT model. This result is reasonable since EFTs are constructed such that they are as predictive as possible at the chosen physical scales of interest. An analysis of the RGE improved potential for a range of parameters of the SM is presented in~\cite{Masina:2024ybn}. The detailed investigation of the parameter space of the SM (and its possible supersymmetric or GUT extensions
at the inflationary or higher energy scales)  to ensure alignment with our model will be the subject of future work.

\end{document}